\documentclass[utf8, nocrop]{FrontiersinHarvardArxiv} 
\usepackage{url,hyperref,lineno,microtype}
\usepackage{subcaption}
\usepackage[onehalfspacing]{setspace}
\usepackage{orcidlink}
\usepackage[figurename=Fig.,tablename=Tab.]{caption}
\usepackage{siunitx}
\usepackage{multirow}
\usepackage{multicol}
\usepackage{booktabs}
\usepackage{amsmath}
\usepackage{color}
\usepackage{pdfpages}
\usepackage{latexml}

\def\N{\mathbb{N}}
\def\R{\mathbb{R}}
\def\B{\mathbb{B}}

\nolinenumbers

\def\keyFont{\fontsize{8}{11}\helveticabold }
\def\firstAuthorLast{Witt, A. {et~al.}} 
\iflatexml
\def\Authors{Arthur Witt\,$^{1,*}$\, \href{https://orcid.org/0000-0003-1180-1172}{[ORCID:~0000-0003-1180-1172]}, Jangho Kim\,$^{2,*}$\, \href{https://orcid.org/0000-0002-4670-0390}{[ORCID:~0000-0002-4670-0390]},\\ Christopher Körber\,$^{3}$\, \href{https://orcid.org/0000-0002-9271-8022}{[ORCID:~0000-0002-9271-8022]}, and Thomas Luu\,$^{2}$\, \href{https://orcid.org/0000-0002-1119-8978}{[ORCID:~0000-0002-1119-8978]}}
\else
\def\Authors{Arthur Witt\,$^{1,*}$\,\orcidlink{0000-0003-1180-1172}, Jangho Kim\,$^{2,*}$\,\orcidlink{0000-0002-4670-0390}, Christopher Körber\,$^{3}$\,\orcidlink{0000-0002-9271-8022}, 
and Thomas Luu\,$^{2}$\,\orcidlink{0000-0002-1119-8978}}
\fi

\def\Address{$^{1}$Institute of Communication Networks and Computer Engineering, University of Stuttgart, Stuttgart, Germany\\
$^{2}$Institute for Advanced Simulation (IAS-4) \& JARA HPC, Forschungszentrum Jülich, Jülich, Germany\\
$^{3}$Fraunhofer Research Institution for Energy Infrastructures and Geothermal Systems IEG, Fraunhofer IEG, Bochum, Germany}

\def\corrAuthor{Arthur Witt}
\def\corrEmail{arthur.witt@ieee.org}

\def\diag{\mathrm{diag}}
\def\ve#1{\mathlette{\boldmath}{#1}}
\def\mathlette#1#2{{\mathchoice{\mbox{#1$\displaystyle #2$}}%
        {\mbox{#1$\textstyle #2$}}%
        {\mbox{#1$\scriptstyle #2$}}%
        {\mbox{#1$\scriptscriptstyle #2$}}}}

\def\T{\mathrm{T}}
\def\argmin#1{\mathrm{arg}\underset{#1}{\mathrm{min}}}
\def\min#1{\underset{#1}{\mathrm{min}}}

\graphicspath{{fig_jpeg}}
\begin{document}
\onecolumn
\firstpage{1}

\title {ILP-based Resource Optimization Realized by Quantum Annealing for Optical Wide-area Communication Networks—A Framework for Solving Combinatorial Problems of a Real-world Application by Quantum Annealing} 

\iflatexml
\author[\firstAuthorLast ]{\Authors\\\Address\\\textbf{Correspondence:}~arthur.witt@ieee.org or j.kim@fz-juelich.de}
\else
\author[\firstAuthorLast ]{\Authors}
\fi 
\iflatexml
\else
\address{} 
\correspondance{} 


\extraAuth{Jangho Kim\\j.kim@fz-juelich.de}
\fi

\maketitle

\begin{abstract}
Resource allocation of wide-area internet networks is inherently a combinatorial optimization problem that if solved quickly, could provide near real-time adaptive control of internet-protocol traffic ensuring increased network efficacy and robustness, while minimizing energy requirements coming from power-hungry transceivers.  In recent works we demonstrated how such a problem could be cast as a quadratic unconstrained binary optimization (QUBO) problem that can be embedded onto the D-Wave Advantage\texttrademark\  quantum annealer system, demonstrating proof of principle.  Our initial studies left open the possibility for improvement of D-Wave solutions via judicious choices of system run parameters.  Here we report on our investigations for optimizing these system parameters, and how we incorporate machine learning (ML) techniques to further improve on the quality of solutions. In particular, we use the Hamming distance to investigate correlations between various system-run parameters and solution vectors.  We then apply a decision tree neural network (NN) to learn these correlations, with the goal of using the neural network to provide further guesses to solution vectors.  We successfully implement this NN in a simple integer linear programming (ILP) example, demonstrating how the NN can fully map out the solution space that was not captured by D-Wave.  We find, however, for the 3-node network problem the NN is not able to enhance the quality of space of solutions.

\iflatexml
\else
\tiny
 \keyFont{ \section{Keywords:} 
     discrete optimization, 
     integer linear program,
     machine learning, 
     quantum annealing, 
     quantum computing, 
     resource allocation,
     wide-area networks} 
\fi
\end{abstract}

\section{Introduction}

Quantum computing is a cutting-edge technology that has gained significant relevance during the last decades. Algorithms for searching and optimization are currently studied intensively on quantum computers as they hold the potential for solving problems with non-polynomial (NP) complexity very efficiently. Nowadays, quantum computers have reached a scale that allows for the solution of non-trivial problems which have real-world applications. 

One example is the energy-aware resource allocation of wide-area networks~\cite{EATE}. In these cases, one can consider the resource allocation as an optimization problem and introduce it as a relevant application of quantum computing, and in particular quantum annealing (QA), as it inherently provides a certain failure tolerance with self-healing capability. Further, the problem has NP complexity and requires frequent solutions for just-in-time adaptation of the network. If solutions are generated quickly, say on the order of seconds, a revolution in network operation with increased network efficiency might be possible since current solutions obtained from classical and/or heuristic algorithms require 15 minutes or more for time-to-solution as shown in \cite{ILPTornatore} and \cite{FellerSA}. In previous studies, \cite{Witt:2022lsx},  we have demonstrated how this resource allocation problem can be formulated, based on an integer linear program (ILP) model, as a quadratic unconstrained binary problem (QUBO) which can then be embedded onto a quantum annealer.

In our initial studies we used the D-Wave Advantage\texttrademark\ system (\texttt{JUPSI}) at the Forschungszentrum J\"ulich to perform the quantum annealing. As part of the solution process, the QUBO problem was embedded onto the quantum qubits prior to performing the quantum annealing.  This entailed mapping the problem onto a network of logical qubits, whereby each logical qubit consists of a constellation, or `chain', of physical qubits.  This mapping ensures the requisite `connectivity' of the logical qubits as dictated by the QUBO problem.   We discovered that the network optimizing approach was greatly limited by this embedding process.  For example, the optimization problem of a network with 3 nodes can be described as a QUBO with approximately 100 binary variables (logical qubits).  Even though this 3-node problem is quite small, the required amount of physical qubits was in the range of 500 qubits, representing already roughly 10\% of the  physical qubits available in the D-Wave Advantage\texttrademark\ system.  With the current embedding process that we employed at the time, a simulation of a 15-node problem, corresponding to a real-world network, would require a quantum annealer with approximately 50,000 physical qubits, which is an order of magnitude larger then current systems.  We note that the embedding process is not unique.

Our initial studies also had limited scope in system run parameters, such as annealing time and profile of the annealing process, penalty factor of the QUBO matrix, and chain strength between physical qubits constituting logical qubits.  Our choice of run parameters were constrained mainly to system default values, with little exploration on the dependence of quality of solutions on these run parameters.  Therefore there is potential room for increasing the efficiency of the quantum annealing process (which would result in better quality and more \emph{feasible} solutions) by judicious choice of optimized run parameters.

In this paper we address some of these issues by studying the process of annealing with the aim to optimize the parameters for the quantum annealing procedure. We introduce solution quality metrics for evaluation purposes. Of particular import is the Hamming distance metric, which rates the distance between the ideal and obtained solution vector in binary space.   By using D-Wave solutions in conjunction with the Hamming distance to optimal solution,  we empirically determine correlations between various run parameters and the quality of solution. 
These correlations guide us in determining optimized run parameters for the system in question, with the hope that the same optimized run parameters can be applied to similar, but larger, systems.  Furthermore, we apply a decision tree neural network (NN) to learn these correlations, after which we use the NN to `guess' improved solutions.   This NN represents a machine learning (ML) approach that we couple with D-Wave generated solutions that aims at providing better quality solutions, and represents an example of a hybrid classical (ML)/quantum (QA) procedure for solving the combinatorial optimization problem.

Our paper is organized as follows.  In the following section we give a cursory description of ILPs in general, the used method to solve ILPs on quantum annealer, and two ILPs that we examined in our study.  We then introduce in \autoref{sec:solution_quality} a Hamming distance metric, and demonstrate how it is used to derive correlations between quality of solutions and various system run parameters.   Such correlations will  be `learned'  by our decision tree NN, which we describe in detail in \autoref{sec:NN}.  In \autoref{sec:results} we present our findings.  We first concentrate on a simple ILP problem, demonstrating that our hybrid classical/quantum procedure does indeed result in new feasible solutions while at the same time providing guidance on optimized run parameters.  We then apply the formalism to the 3-node network problem mentioned above, where here we see limited improvement in solutions, all of which unfortunately are nowhere near the optimal solution.  In \autoref{sec:conclusion} we discuss our findings and recapitulate.  We comment on possible future directions of investigation.

\section{Materials and Methods}

\subsection{The Concept of Integer Linear Programs (ILPs)}
The investigations performed in our work fall under the class of discrete optimization problems, meaning variables $\ve x$ to be optimized take on only discrete values.  Such problems can be cast succinctly as an integer linear program (ILP), 
where certain constraints, given as a set of linear (in-)equation, have to be satisfied while minimizing a linear function. An ILP can be defined in its canonical form by
\iflatexml
\begin{align}
	\mathrm{objective}&\quad &\ve x_0 = \argmin{\ve x}\{\ve c^\T\ve x\}\label{eq:ilp_small_obj}\\
	\mathrm{constraints}&\quad &\ve A \ve x+\ve b \leq \ve 0\label{eq:ilp_small_const}\\
	\mathrm{variables}&\quad &\ve x \in \N^n,\, x_i \geq 0\label{eq:ilp_small_var}\\
	\mathrm{constants}&\quad &\ve c \in \R^n, \ve b \in \R^m, \ve A \in \R^{m\times n}\label{eq:ilp_constants}\,.
\end{align}
\else
\begin{eqnarray}
    \text{objective} &&\ve x_0 = \argmin{\ve x}\{\ve c^\T\ve x\}\label{eq:ilp_small_obj}\\
    \text{constraints} &&\ve A \ve x+\ve b \leq \ve 0\label{eq:ilp_small_const}\\
    \text{variables} &&\ve x \in \N^n,\, x_i \geq 0\label{eq:ilp_small_var}\\
    \text{constants} &&\ve c \in \R^n, \ve b \in \R^m, \ve A \in \R^{m\times n}\label{eq:ilp_constants}\,.
\end{eqnarray}
\fi
The ILP's objective function can be seen as a loss function and is defined in \eqref{eq:ilp_small_obj} with a vector of cost terms $\ve c$ weighting the variable vector $\ve x$. Matrix $\ve A$ and vector $\ve b$ parameterize the linear equations that represent the inequality constraints~\eqref{eq:ilp_small_const}. They can be reshaped to equality constraints, $\ve A\ve x+\ve b + \ve s= \ve 0$, by introducing \emph{slack} variables $\ve s\in \R, s_j \geq 0$. This is a typical step within the classical ILP-solving algorithm \emph{simplex}, see~\cite{simplex}.  
We use the convention, that $\R$ are real-valued numbers, $\N$ natural numbers inclusive zero, and $\B$ binary numbers. 

Such problems are well known to be non-polynomial (NP)-hard in general. According to \cite{complexity_ilp} and \cite{Karp1972}, linear programs are a rare class of problems in NP that resists the classification as NP-complete or polynomial-solvable problems. \cite{Lenstra} argued, that mixed-integer linear programs with fixed number of variables are solvable in polynomial-time. In contrary, \cite{nguyen2017computational} present integer programs that are NP-complete, even for fixed number of variables. Their work further shows, that some integer programs are solvable in polynomial-time. We can conclude, that bounded integer linear programs are NP-complete and are solvable within polynomial time in few cases. 

\subsection{Solving Integer Linear Programs on Quantum Annealer}
\label{sec:ilp_qa_qubo}

Quantum annealers are well suited for investigating ILP problems.  However, an additional modification to the ILP problem is required prior to embedding the problem on the quantum qubits.  Here the constraints are included into the cost function (to be minimized) by introduction of penalty weight $p$.  In so doing, the original ILP problem with constraints is recast into quadratic form without constraints,
\begin{equation}
    \left.
    \begin{aligned}
        \ve{x}_0
        &=
        \argmin{\ve{x}}\left\{\ve{c}^\top \ve{x} \right\}\\
        \ve A \ve{x}+\ve{b}
        &
        \leq \ve{0}
        \\
        \ve{x}
        &\in\N^n \geq 0
    \end{aligned}
    \right\}
    \longleftrightarrow
    \left\{
    \begin{aligned}
        \ve{q}_0 &= \argmin{\ve q}\left\{\ve{q}^\top \ve Q \ve q + C \right\}\\
        \ve q&\in \{0, 1\}^k
    \end{aligned}
    \right.
    \,.
\end{equation}
Here $\ve A$, $\ve b$ and $\ve c$ are problem specific parameters as introduced in the previous section.
It is useful to classify solution vectors $\ve{x}$ into two categories: \textbf{feasible} solutions which fulfill the constraints and \textbf{unfeasible} solutions which violate the constraints.
While a feasible solution to the ILP is not necessarily an optimal solution, an unfeasible solution hypothetically can have a smaller objective value than the optimal feasible solution.

The problem is mapped to the quadratic unconstrained binary optimization (QUBO) by definition of matrix $\ve Q$ that includes a penalty factor $p$ and a constant $C$.
The inequality can be expressed by an equation and another minimization over a slack variable $\ve{s}$ incorporated into the bit vector $\ve{q}$.
The QUBO objective function minimizes both the ILP objective function plus another objective function representing the constraints.
The penalty term expresses the relative weight between both (ILP objective and constraint) objective functions, see \cite{Chang:2020iwh} and~\cite{Witt:2022lsx} for a more detailed description.  Finding the solution set $\ve{q}_0$ that provides the absolute minimum of $\ve{q}^\top\ve{Q}\ve{q}$ is equivalent to solving the original ILP problem with solution vector $\ve{x}_0$.

The D-Wave Advantage\texttrademark\  system is adapted to solving the Ising spin system that represents an array of binary spins with interactions between spins $\ve{\sigma}$ giving by some connectivity matrix $\ve{J}$ and external magnetic field $\ve{h}$.  Our QUBO matrix can easily be rewritten using $\ve{J}$ and $\ve{h}$ without any loss of generality, 
\begin{equation}
    \ve{q}^\top \ve{Q} \ve{q}+ C
    \Leftrightarrow
    \ve{\sigma}^\top \ve{J} \ve{\sigma}
    +
    \ve{h}^\top \ve{\sigma}
    +
    g
    \text{\quad with\quad }
    \begin{cases}
        \ve{J} &= \frac{1}{4} \ve Q_0\\
        \ve{h} &= \frac{1}{2} \ve{\hat{q}} + \frac{1}{2} \ve Q_0\ve{1} \\
        g &= \frac{1}{4}\ve{1}^\top \ve Q_0 \ve 1 + \frac{1}{2} \ve 1^\top \ve{\hat{q}}+ C \\
    \end{cases}
    \,,
\end{equation}
with  $\ve Q_0 = \ve  Q - \diag\{\ve{\hat{q}}\}$, $\ve{\hat{q}} = \diag^{-1}\{\ve Q\}$, and g some constant.
The problem is now well suited for the D-Wave machine.
In \cite{Chang:2020iwh} and \cite{Witt:2022lsx}, we demonstrated proof of principle that such a problem can be solved on a quantum annealer.

\subsection{Investigated Integer Linear Programs}
In this work we have investigated multiple ILPs, two of which we define explicitly here.  
The details of the remaining ILPs we considered are described in our accompanying Supplementary Material.
The first ILP optimizes the selection of two integer variables under some constraints. It provides a test case where all possible solutions can be studied with the approach of brute force sampling, i.e., it provides a well-suited setup for benchmarking. The second ILP describes a realistic network resource optimization as studied in \cite{Witt:2022lsx}. As possible solutions are representable as binary vectors with more than 60 variables, a brute force sampling is not applicable within reasonable time for this case.

\subsubsection{Trivial ILP}
\label{sec:trivial_ilp}
\begin{figure}[h]
	\centering
	\includegraphics[width=0.8\textwidth]{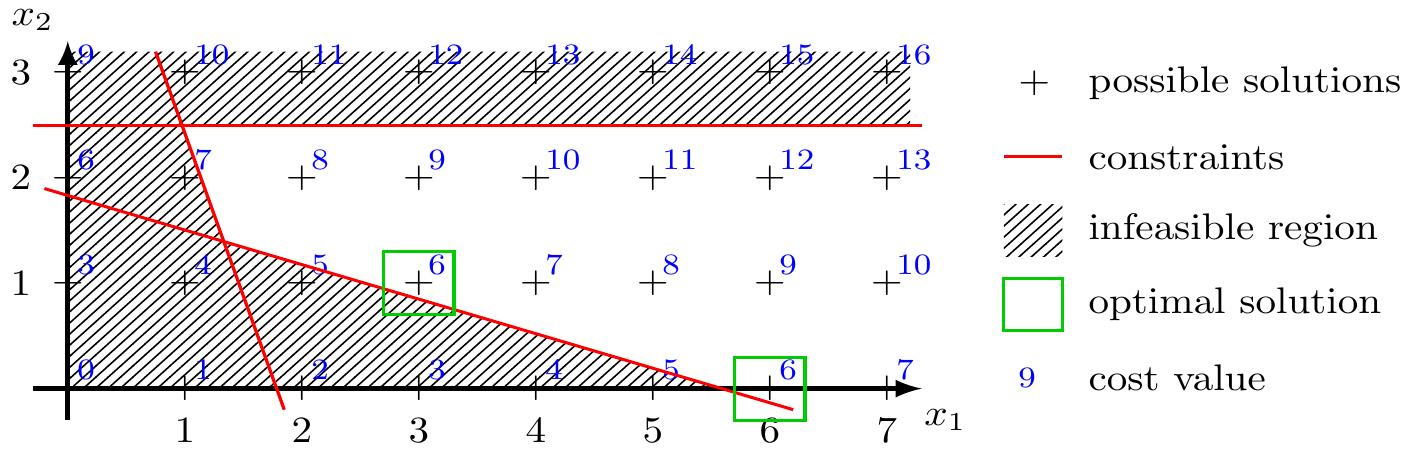}
	\caption{Graphical interpretation of the trivial ILP problem. The drawn constraint lines are slightly shifted for visualization purposes without falsifying the feasible region of integer values.}
	\label{fig:ilp}
\end{figure}

Based on expressions \eqref{eq:ilp_small_obj} to \eqref{eq:ilp_constants}, we can define a particular ILP problem by 
\begin{equation}
    \ve A=\begin{bmatrix}
        -1/3 &-1\\
        -3 & -1\\
        0 & 1\\
    \end{bmatrix}\,,\quad
    \ve b=\begin{bmatrix}
        2\\6\\-2
    \end{bmatrix}\,,\quad
    \ve c =
    \begin{bmatrix}
        1\\3
    \end{bmatrix}\,,\quad
    \ve x\in \N^2\,,\quad
    \ve s\in \N^3\,.
\end{equation}
A graphical interpretation of this ILP is depicted in \autoref{fig:ilp}. We can easily obtain the optimal solution vectors,
\begin{equation}
    \ve x_0 = \begin{bmatrix}
        3\\1
    \end{bmatrix}\, \text{or}\, \begin{bmatrix}
        6\\0
    \end{bmatrix}
    \,,
\end{equation}
and the optimal cost value $\ve c^\T\ve x_0=6$ from this graph. This problem is an explicit example of an ILP that can have  more than one optimal solution, which in turn can cause misleading results in benchmarking experiments. In general, ILPs can have zero, one or more solutions. We restrict the values in $\ve x$ to $x_i\in\{0,1,2,3\}, \forall i\in\{1,2\}$, such that the ILP is uniquely solvable.

For binary representation of integer values, we will use 2 bits for each variable in $\ve x$ and 3 bits for each element in $\ve s$. Then $\ve q$ is the binary search vector to be optimized. According to \cite{Chang:2020iwh}, this mapping with integer mapping matrix $\ve Z$ can be described by

\begin{equation}
\setcounter{MaxMatrixCols}{15}
\begin{bmatrix}
    \ve x\\--\\
    \ve s\\
\end{bmatrix}
=
\begin{bmatrix}
    x_1\\x_2\\--\\s_1\\s_2\\s_3
\end{bmatrix}
=   
\underset{\ve Z}{\underbrace{
    \begin{bmatrix}
        2&1&0&0&|&0&0&0&0&0&0&0&0&0\\
        0&0&2&1&|&0&0&0&0&0&0&0&0&0\\
        0&0&0&0&|&4&2&1&0&0&0&0&0&0\\
        0&0&0&0&|&0&0&0&4&2&1&0&0&0\\
        0&0&0&0&|&0&0&0&0&0&0&4&2&1\\
    \end{bmatrix}
}}
\underset{\ve q}{\underbrace{
\begin{bmatrix}
q_1\\q_2\\\vdots\\q_{13}
\end{bmatrix}}}\,.
\end{equation}

In Tab.~S1 of our Supplementary Material we enumerate other trivial ILPs that we have investigated.  These ILPs encompass a range of optimal solutions, parameters, and dimensions.

\subsubsection{Network Resource Allocation Problem}

\begin{figure}[h]
	\centering
	\includegraphics[width=0.8\columnwidth]{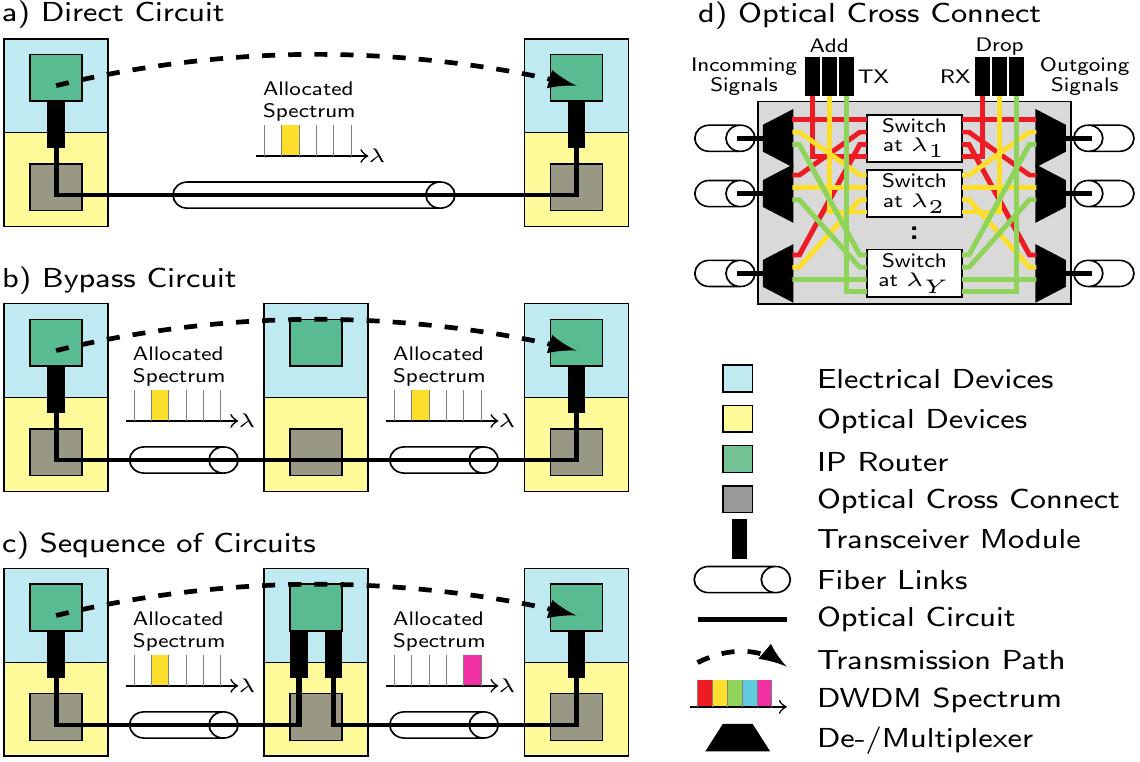}
	\caption{a--c) Various ways of realizing transmission paths in wide-area networks with optical DWDM layer and d) architecture of a OXC as introduced in \cite{Witt:2022lsx}.}
	\label{fig:tp_realization}
\end{figure}

Optical wide-area networks consist of nodes that are linked by optical fiber systems in form of a meshed topology. Nodes $v\in V$ are two layered. They are equipped with electrical IP routers in the upper layer and optical cross connects (OXCs) in the lower layer. Traffic from connected networks that traverses the wide area network is ``handed over" at the IP layer. Signal transitions between layers inside the WAN are performed with optical bidirectional transceivers, that are configured for unidirectional use as required. 
Optical transceivers generate optical signals with a bandwidth of \si{50}{GHz} at various center frequencies. A finite number of signals can be combined in a dense wavelength division multiplexing (DWDM) scheme on a particular optical fiber link. This schemes are specified according to \cite{itu_t_dwdm_grid}. Thus, usable frequency bands in the optical region, typically referred as wavelengths, are uniquely defined. Optical cross connects enable wavelength-selective forwarding and redirection of optical signals between connected fibers. Fiber links are realized by a sequence of fiber spans and fiber amplifiers and provide a hardware-wise connection between nodes according to the networks topology. The maximal reach of optical signals depends on the signal configuration (specified by modulation schemes, and used forward error correction, etc.) and the transceiver type itself. As an example, a tunable coherent  transceiver\footnote{100G/200G Tunable Coherent CFP2-DCO Transceiver:\\ \url{https://www.fs.com/de-en/products/120128.html?attribute=5320&id=297112}} 
achieves an optical reach of 1000 km at a rate of 100 GBit/s. Typically, optical transmission paths are organized as a sequence of transmission sections $c$ with at least one section to enable a end-to-end data transfer. Transmission sections are abstract links in the lower layer that provide optical transparent transmission on multiple wavelength. Their spanning distance is limited by the optical reach of the driving transceivers. \autoref{fig:tp_realization} illustrates how transmission paths in wide-area networks can be realized.

Energy-aware traffic engineering can be seen as a major task for economic network operation. Therefore, network resources like transceivers and wavelengths on fiber links have to be allocated to assign the required capacity to a transmission section  $c$. Assuming that the network is operated as single rate system, i.e. all transceivers have the same signal rate, e.g. $\xi=\SI{100}{GBit/s}$, capacities at a transmission section $c$ can be scaled if multiple transceivers, enumerated by $w_c$, are activated. Thus, the transmission section's capacity is $w_c\xi$. A unidirectional traffic demand $d$ represents a connectivity request between two network nodes. We assume, that a demand exist for all disjunct node pairs ($u,v$) with $u\neq v$ and $u,v \in V$. The network has to provide appropriate transmission paths, i.e. routes through the network topology along a sequence of transmission sections, to enable the transport of the demand's traffic with volume $h_d$. We prepare a network-specific collection $T$ of possible transmission path realization prior the optimization, whereas possible transmission path realizations per demand $d$ are defined as $t_d\in T_d \subset T$, see \cite{Witt:2022lsx} Sec.~II-C. The economic resource allocation within WDM networks is a discrete and combinatorial optimization problem.

\cite{Witt:2022lsx} devised an integer linear program (ILP) based on \cite{IKR2} to address the energy-aware resource allocation problem within wide-area networks and prepared the ILP according to the ILP-to-QUBO mapping formalism as presented in \cite{Chang:2020iwh} and delineated in \autoref{sec:ilp_qa_qubo}. They further studied the solvability of this ILP, prepared in QUBO form, on the D-Wave Advantage\texttrademark, a state-of-the-art quantum annealer with over 5000 qubits. Since the current work focuses on improvement methods within the algorithmic part and not on the application itself, we refer to \cite{Witt:2022lsx} for a more in-depth explanation and interpretation of the ILP. In the following, we recapitulate the ILP briefly. Parameters and variables are given in \autoref{tab:ilp_param}. Traffic volumes $h_d$ per demand $d$, that are varying over time, are held constant during the optimization and will be updated frequently in a real scenario. The equality constraint~\eqref{eq:const_demand} enforces that a demand is routed on exactly one transmission path. Constraint~\eqref{eq:const_circuits} combines traffic flows per transmission section as selected in \eqref{eq:const_demand} and reserves the required capacity in terms of optical channels $w_c$. 
Constraint~\eqref{eq:const_nodes} activates installed transceivers to drive the transmission sections. 
Minimizing the number of optical channels $w_c$ as defined in objective~\eqref{eq:objective}, reduces the total amount of active transceivers as well. This enables a energy-aware network operation. 

\noindent\textit{Constraints}:
\newlength{\negspace}
\setlength\negspace{0mm}
\begin{eqnarray}
    \hspace{-5mm}\sum_{t_d \in T_d} g_{t_d} = 1 \hspace{\negspace}&&\forall d \in D\hspace{5mm}
    \label{eq:const_demand}\\
    -w_c +
    \sum_{d \in D}
    \sum_{t_d \in T_d}
    \rho_{c,t_d} \cdot\frac{h_d}{\xi} \cdot g_{t_d} \le 0 \hspace{\negspace}&&\forall c \in C
    \label{eq:const_circuits}\\
    \sum_{c \in C} w_c \cdot   \varphi_{v,c} \le \eta_v  \hspace{\negspace}&&\forall v \in V
    \label{eq:const_nodes}
\end{eqnarray}
\textit{Objective:}
\begin{eqnarray}
    \sum_{c \in C} w_c\ \rightarrow\ \min.\label{eq:objective}
\end{eqnarray}

The network under test is a fully-connected 3 node network, e.g. the topology has a triangular shape with two short edges of \SI{300}{km} length and a long edge spanning a \SI{424}{km} distance. Each network edge is realized by two fiber links to realize bidirectional transmission. Traffic demand values $h_d$ are taken from a normal distribution with mean  \SI{75}{Gbit/s} and standard deviation  \SI{20}{Gbit/s}. As they represent floating numbers, we discretize them with an accuracy of a=1 (acc. to \cite{Witt:2022lsx} Sec. III-C), i.e. fractions are rounded to `x.0' or `x.5'. We set the number of installed transceivers per node to $\eta_v=15$ and the maximal number of parallel optical signals per transmission path to $\omega_{c,\text{max}}=3$. The parameter $\omega_{c,\text{max}}$ influences the QUBO's matrix sizes as described in \cite{Witt:2022lsx} Sec. III-C. The parameters $\rho_{c,t_d}$ and $\varphi_{v,c}$ represent the connectivity described by the topology. They are predefined together with the transmission path realization sets $T_d$. 
The boolean selector variable $g_{t_d}$, indicating the selection of a predefined transmission path realization $t_d$ for demand $d$, and the number of parallel optical signals per transmission path $\omega_c$ are determined during the optimization.

\begin{table}[t]
	\centering
	\caption{List of parameters used in the ILP for network optimization.}
	\label{tab:ilp_param}
	\begin{tabular}{llp{11.5cm}}
		\toprule
		\multicolumn{2}{l}{Parameter}&Interpretation\\
		\midrule
		\multirow{6}{*}{Constants}&$\xi \in \R$& data rate of a single transceiver\\
		&$\eta_v \in \N$ & amount of transceivers installed at node $v$\\
		&$\rho_{c,t_d} \in \B$& indicates whether the transmission section $c$ is part of the demand-specific transmission path realization $t_d$\\
		&$\varphi_{v,c} \in \B$& indicates whether transmission section $c$ is connected to node $v$\\
		&$h_d\in \mathbb{R}$& traffic volume of demand $d$\\
		\midrule
		\multirow{3}{*}{Variables}&$g_{t_d}\in\B$&path selector is 1 if a transmission path for demand $d$ is realized by circuit configuration $t_d\in T_d$\\
		&$\omega_c\in\N$&amount of active transceivers, driving a transmission section $c$\\
		\bottomrule
	\end{tabular} 
\end{table}

\subsection{Correlations Between Solution Metrics and System Run Parameters}
\label{sec:solution_quality}

With the intent to minimize the objective function, D-Wave provides a distribution of solutions, all of which are not equally important nor of equal quality.
The setup of the ILP scenario (penalty term, float variable solution, integer sizes) and the QUBO (sparsity-affecting transformations, embedding, chain strength) parameterize the problem.
The annealing procedure (annealing schedule, spin transformations, thermalization/decorrelation pauses) can also have significant influence on the obtained distribution of solutions. Studies like \cite{Willsch_2022} show that a proper parameter selection in terms of annealing schedule and embedding variants can change the situation significantly. 
Furthermore, the effect of thermalization in the context of quantum annealing processes can have an impact, as was shown in \cite{term_annealing_2013}.
Ideally, the solutions to the problem should not be affected by the choices for these meta parameters. 
Still, we selected the range of parameters to be tested using our experience garnered from our previous study, see \cite{Witt:2022lsx}.

However, as we show in later sections, different combinations of parameters significantly affect the likelihood of obtaining feasible solutions.
Choices for such meta parameters can be highly correlated.
For example, longer (slower) annealing profiles can provide higher probabilities for finding a feasible solution, yet at the cost of generating fewer total number of solutions. 

Our first studies, \cite{Witt:2022lsx}, found that probabilities for finding a minimal feasible solution for the three-node network problem were at the order of $~ 10^{-4}\%$ and below.  This presented a non-trivial task to evaluate the quality of the distribution of solutions when only having a sample sizes of less than $~ 10^6$.
To address this issue, we formulate statistical measures based on the distribution of samples to quantify the quality of our D-Wave setup. Since the optimal solution is, by definition, a feasible solution, we are interested in the rate in which feasible solutions are produced.  We thus consider the feasibility ratio,
\def\rfeas{r_\mathrm{feasible}}
\def\Nfeas{N_\mathrm{feasible}}
\def\Nsamp{N_\mathrm{samples}}
\begin{equation}
    \rfeas=\frac{\Nfeas}{\Nsamp}\,,\label{eq:feasi_ratio}
\end{equation}
that rates the success of finding $\Nfeas$ feasible solutions within a solution set with $\Nsamp$ samples.

Another metric of choice for solutions in the binary search space that we use in our research here is the Hamming distance 
\begin{equation}
    \text{dist}\{\ve{x},\ve{y}\} = \sum_i XOR(x_i,y_i)\ .
\end{equation}
This metric gives the number of flipped bits between an ideal solution $\ve{x}$ obtained by a classical ILP solver like CPLEX or GLPK and a non-ideal solution $\ve{y}$ obtained by the quantum annealer. 
For binary solution vectors the Hamming distance is equivalent to the $L_2$-norm of the difference between $\ve{x}$ and $\ve{y}$.
This metric provides a sense of `distance' between two solution vectors, essentially telling us how many `bit-flips' are required to bring one solution into another.
Ultimately it allows us to perform a direct comparison between particular D-Wave solution vectors and a known desired solution vector.  

Finally, we can train a neural network (NN) on these correlations, with the goal that once trained, we can use the NN to make further guesses on optimal solutions vectors.  We describe our NN in the following section.

\subsection{Machine-learning Approach}
\label{sec:NN}

\begin{figure}[b]
	\centering
	\includegraphics[width=0.76\textwidth]{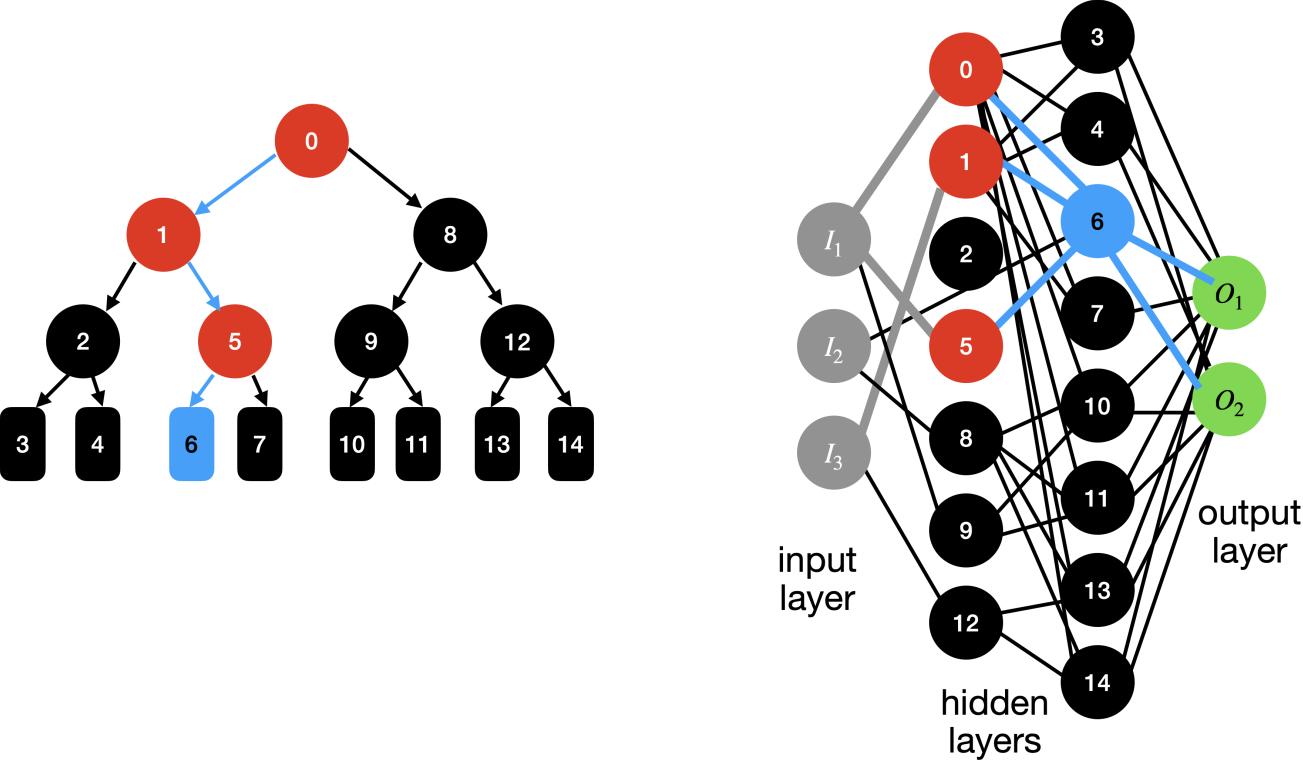}
	\caption{\label{fig:decisionTree} A graphical example of decision tree network (left) and its mapping to a neural network (right).}
\end{figure}

We employ a decision tree (DT) neural network in our investigations.  This NN  is a type of supervised machine 
learning (ML) algorithm that is used typically for regression and classification analysis. 
It is a model that represents a series of 
decisions and their possible consequences in the form of a tree-like structure~\cite{breiman1984classification}. 
Each node in the tree represents a decision, 
and each branch represents a possible outcome or path that can 
be taken based on that decision. 
In Fig.~\ref{fig:decisionTree} we provide a graphical example of a decision tree and its mapping to a neural network.

A major advantage of decision tree NNs is their ease of use, understandability, and interpretability.   
This makes their implementation simple and their application efficient.
Another advantage comes from their inherent robustness to data outliers.
They can even handle missing values in the data.
The data itself can be both categorical and numerical in nature.

However, a potential drawback of DTs is that they can easily overfit the data.
This ultimately means that, though they may be sufficiently \emph{expressive} to explain the trained data, they fail when extrapolating to new, or unseen data.
Thus the NN is limited in its \emph{generalizability}.  
This issue can to a certain degree be mitigated by pruning the tree or using other techniques to reduce the complexity of the model.
In our studies we did not employ such mitigation techniques, and leave such potential studies for later investigations.
We used the \texttt{Scikit-learn} python module~\cite{Pedregosa:2011ork} and its functionalities to implement our DT networks.

\subsection{Construction of Sherrington-Kirkpatrick Graph}
The Sherrington-Kirkpatrick (SK) graph encloses the coupling strength and external fields of a Ising Hamiltonian. As mentioned in \cite{fasthare}, finding the weighted minimal cut in this graph is equivalent to finding the ground state in the Ising Hamiltonian. Further, the Hamiltonian's energy landscape can be explored by exploration of the SK graph's cut space.

The corresponding SK graph of the Ising Hamiltonian $\mathcal{H}(\ve{x})=\ve{h}^\top \ve{x}+\ve{x}^\top\ve{J}\ve{x}$ with $n$ variables $x_i$ can be denoted as $G_H^{SK}=(V,E,w)$ with node set $V$, undirected edges $(i,j)\in E$ and their weights $W$.
The first $n$ nodes of $V$ correspond to the variables $x_i$. A further node is added to $V$ to capture the external fields $\ve{h}$. Set $E$ contains only edges with non-zero weights according to $w_{ij}=J_{ij}+J_{ji}$ for $1\leq i,j\leq n$ and weights $w_{i,n+1}=h_i$. Then, the weighted adjacency matrix $\ve{J}'$ of graph $G_H^{SK}$ with $J_{ij}'=J_{ij}+J_{ji}$, $J_{ji}'=0$ and $J_{i,n+1}'=h_i$ can be used together with $\ve{y}\in\mathbb{S}^{n+1}$ to define the SK Hamiltonian as 
\begin{equation}
    \mathcal{H}^{SK}(\ve{y})=\ve{y}^\top\ve{J}'\ve{y}.
\end{equation}
To apply a weighted minimal-cut approach on the SK graph for graph reduction, a cut is defined by a subset $S \subseteq V$, such that $\langle S,V$\textbackslash$ S\rangle$ contains a set of edges that needs to be cut for separation of $S$ and $V$\textbackslash $X$. With $c(S)=\sum_{(u,v)\in\langle S,V\text{\textbackslash} S\rangle}w_{uv}$, the capacity of the cut, a minimal cut is defined as $S^\ast=mc(G^{SK})=\argmin{S\subseteq V} c(S)$ with the minimal capacity of $MC(G^{SK})=\min{S\subseteq V} c(S)$.

\section{Results}
\label{sec:results}

\subsection{Trivial ILP Problem}
We now provide our findings for our simple ILP problem that we described in \autoref{sec:trivial_ilp}.  Similar results for the other trivial ILPs we considered are found in the Supplementary Material.  Note that this problem is sufficiently small that we can determine all possible feasible solutions via brute force, which includes the optimal solution.  In this case this corresponds to a total number of $\Nfeas=1536$ feasible solutions. The whole solution space contains $\Nsamp=2^{13}=8192$ possible vectors as our binary search vector $\ve q$ has a dimension of 13, see \autoref{sec:trivial_ilp}. Thus, the feasibility ratio \eqref{eq:feasi_ratio} for brute force sampling is $\rfeas=18.75\%$.

\subsubsection{Observations and Correlations}

\begin{figure}[b]
	\includegraphics[width=.495\textwidth]{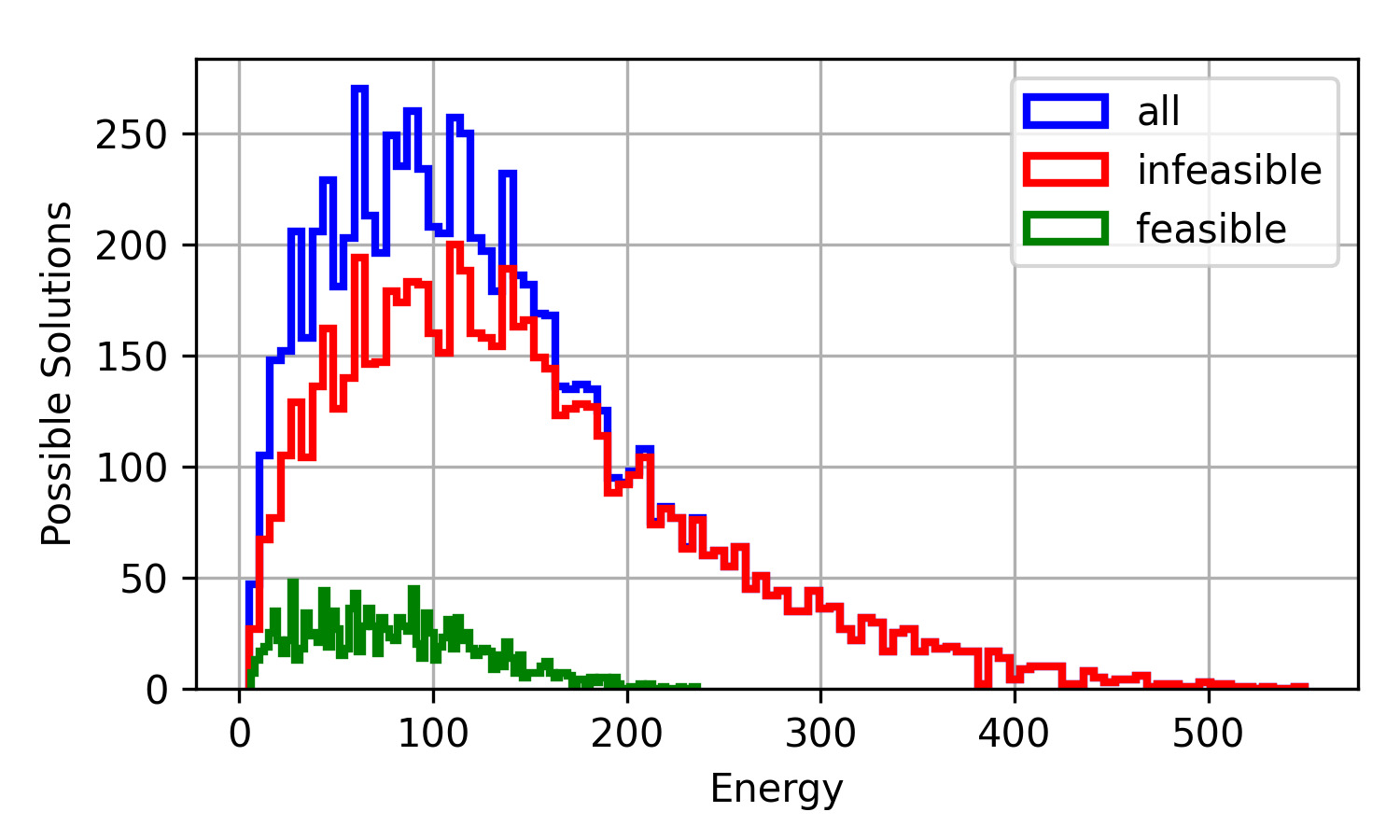} 
	\includegraphics[width=.495\textwidth]{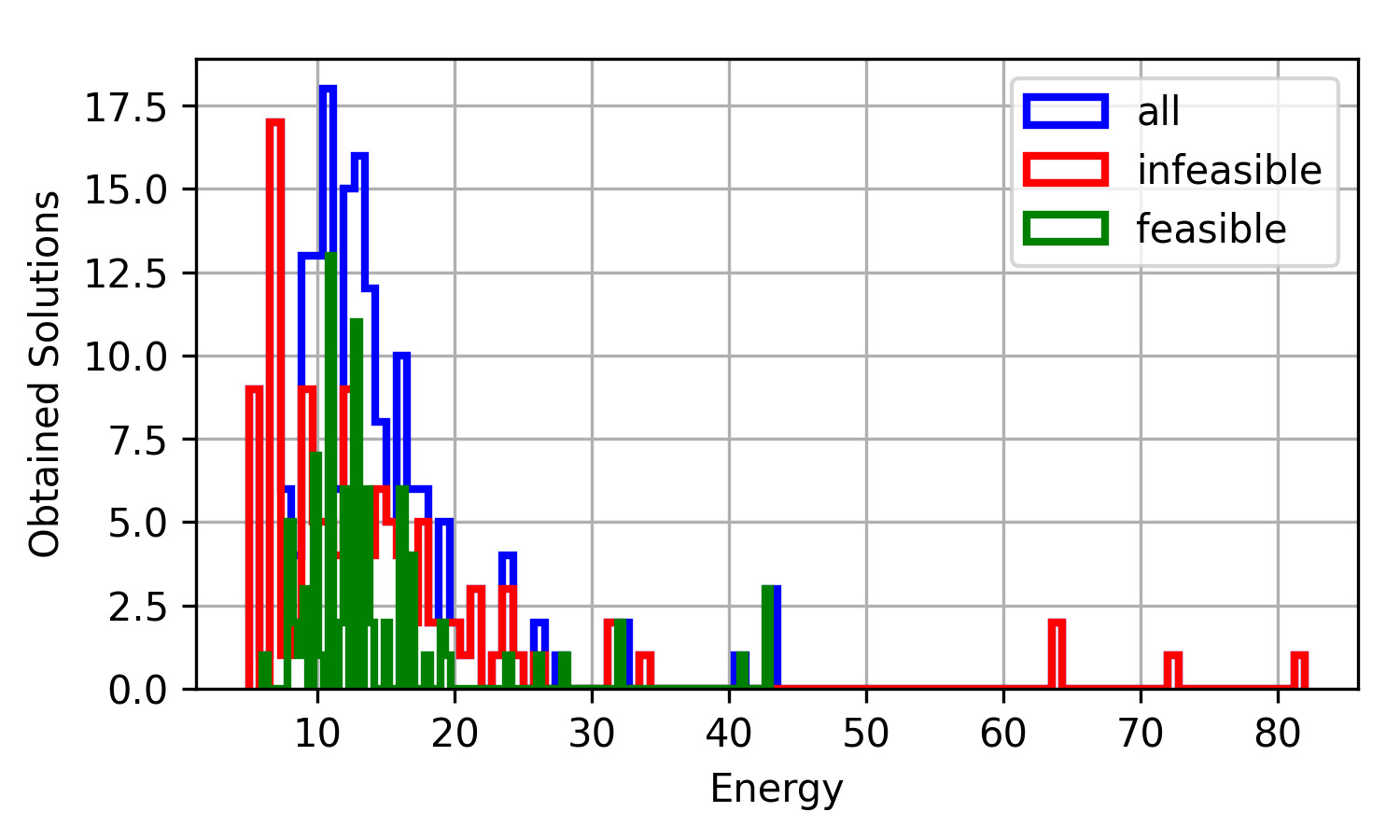}
	\includegraphics[width=.495\textwidth]{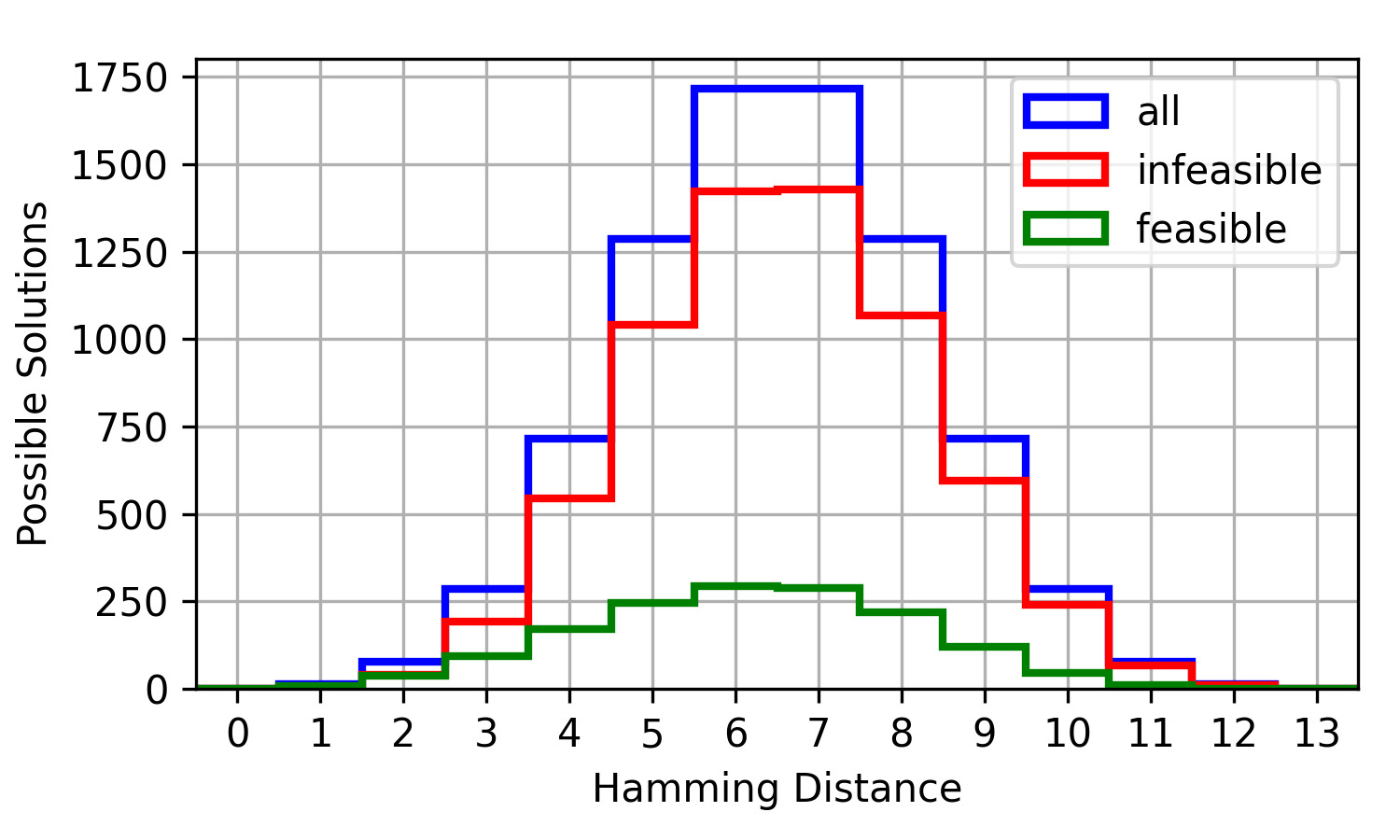}
	\includegraphics[width=.495\textwidth]{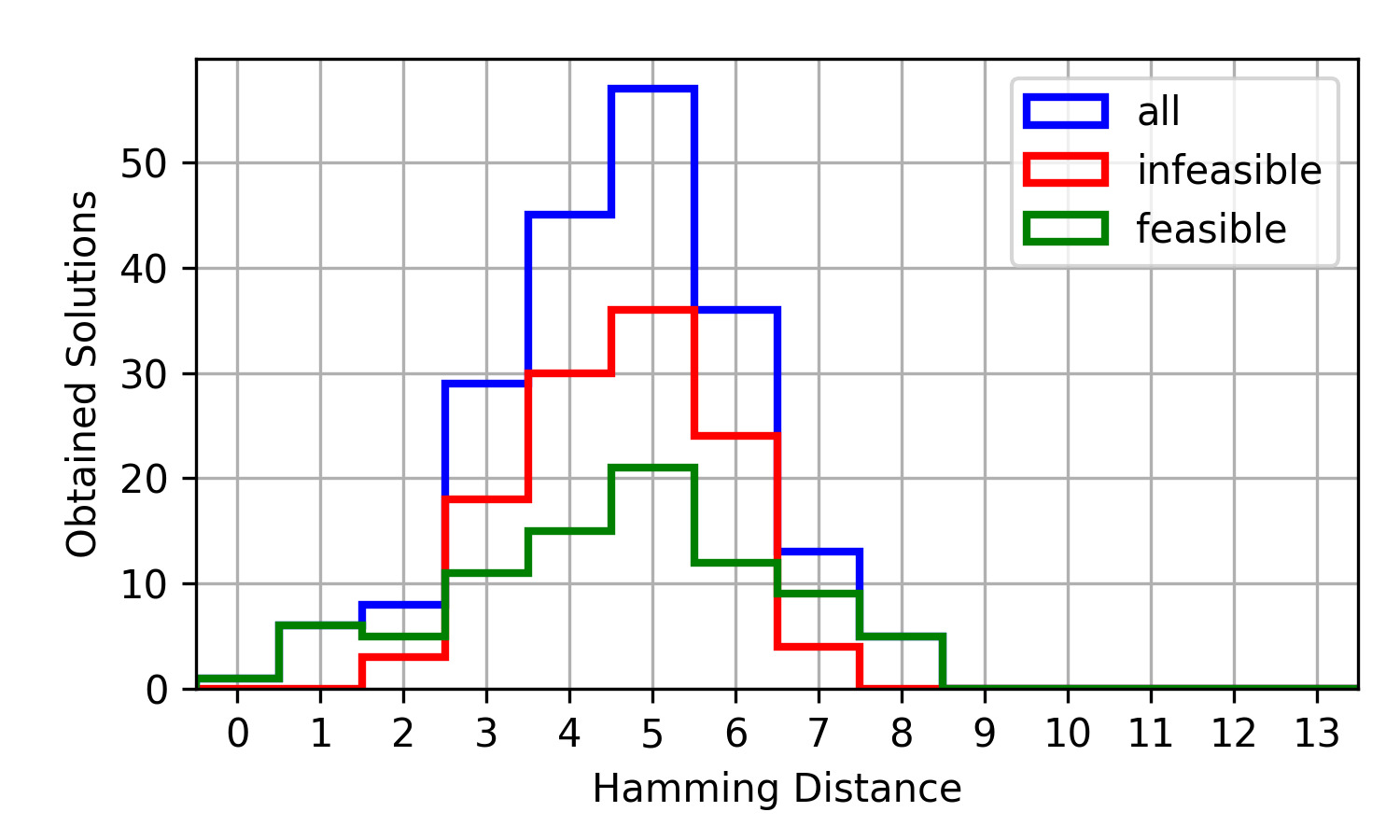}
	\includegraphics[width=.495\textwidth]{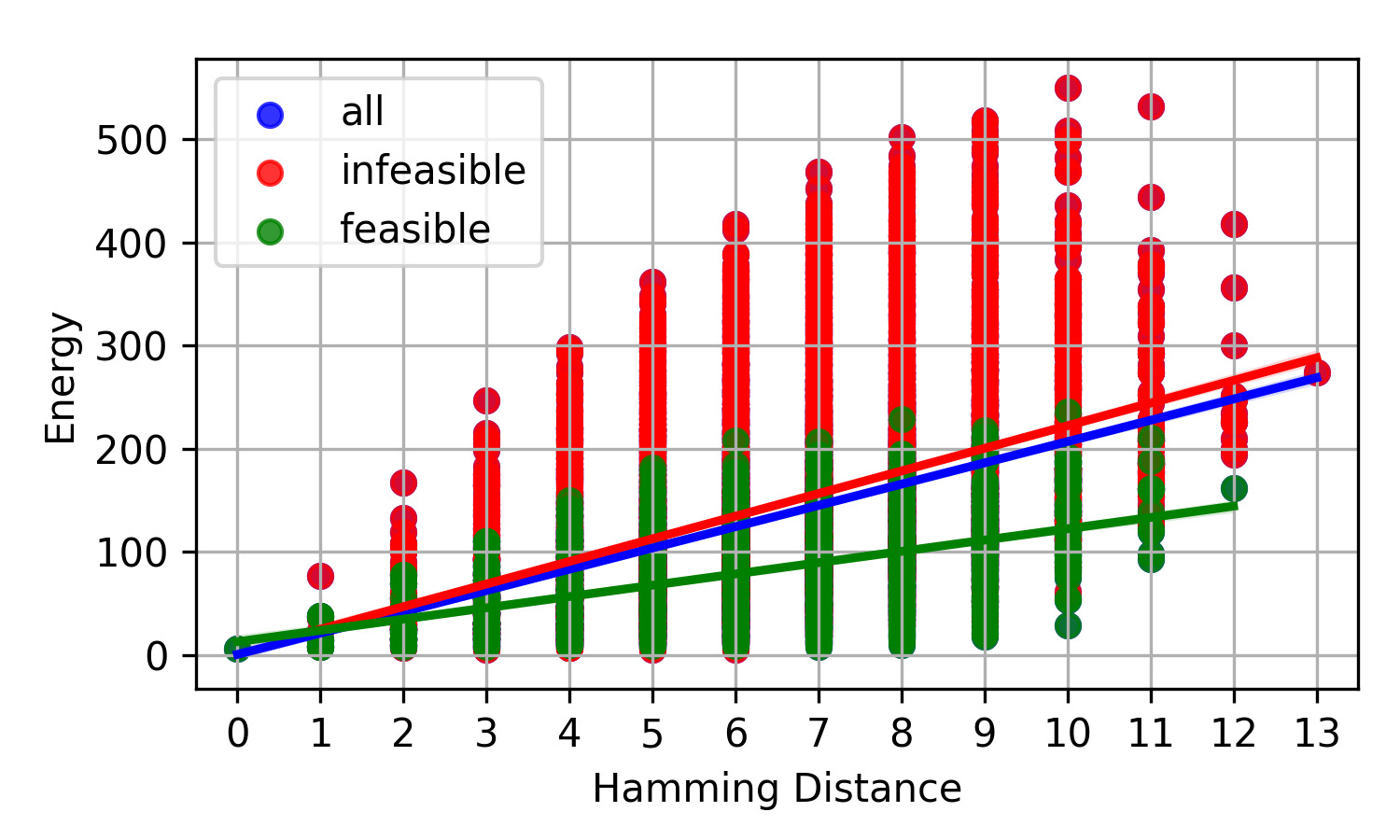}\hspace{1mm}
	\includegraphics[width=.495\textwidth]{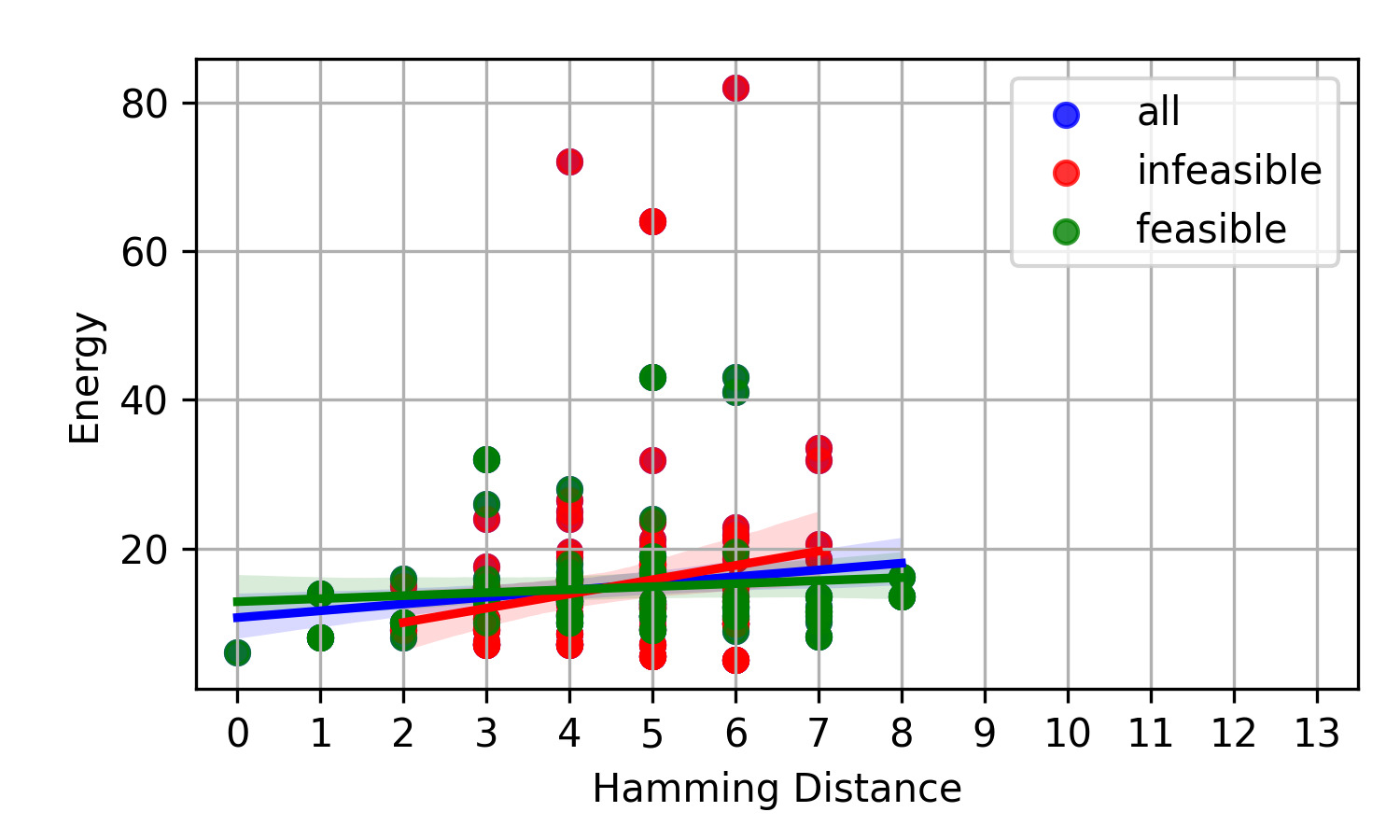}
	\caption{(Left side) brute force sampling to investigate the entire solution space. (Right side) D-Wave sampling with penalty $p=2$ and a set of 200 samples. (Upper row) histogram of solutions sorted by energy values. As solutions  gathered by D-Wave's quantum annealer have only energy values in the lower area compared to the brute force case, x-axis are scaled differently. (Middle row) histogram of solutions over Hamming distance with respect to the best feasible solution vector. (Lower row) scatter plot of solutions with reference to their energy values and the Hamming distance  with respect to the best feasible solution vector. (Blue) solution set under investigation. (Red) infeasible solutions. (Green) feasible solutions.}
	\label{fig:dwave solutions ILP}
\end{figure}

In \autoref{fig:dwave solutions ILP} we show results for brute force sampling (left side) and a run on the D-Wave Advantage\texttrademark\ (right side) using a penalty of $p=2$ and default run parameters. In the D-Wave case, just 200 samples are taken, which is a relative small portion ($\sim 2.4\%$) compared to the complete solution space. We have to remark, that the optimal solution can be found even if the sample set is small. \autoref{fig:dwave solutions ILP} shows the distribution of solutions over energy (upper row) and Hamming distance (middle row) obtained by the mentioned sampling methods and classified by their feasibility demarcated by feasible (green),  infeasible (red), and all (blue) solutions. We can observe, that solutions obtained with D-Wave show low energies and only Hamming distances of up to 8. This indicates, that the aimed optimization takes place and only solutions with mostly good qualities are found by D-Wave quantum annealer. But, we still have to sort solutions by feasibility after sampling as minimizing the energy can not entirely sort out infeasible solutions. 
The lower row of \autoref{fig:dwave solutions ILP} shows the correlations between the solution's energy and their Hamming distance in relation to the best feasible solution. Solutions with small Hamming distances tend to have smaller energy values as observable and indicated by the best fitting curves. We identified, that higher-energy solutions are correlated with increasing Hamming distances to optimal solution as the slope of the fitting curves are non-zero. The energy range for solutions at same Hamming distances is spread widely if the whole search space is considered. Feasible solutions could be found only at the lower energy range. Within the D-Wave sample set, solutions with small Hamming distances are over-proportionally feasible solutions which is indicted by the regression curves.

\begin{figure}[b]
	\centering
	\includegraphics[trim=5mm 0 15mm 5mm, clip, width=0.495\textwidth]{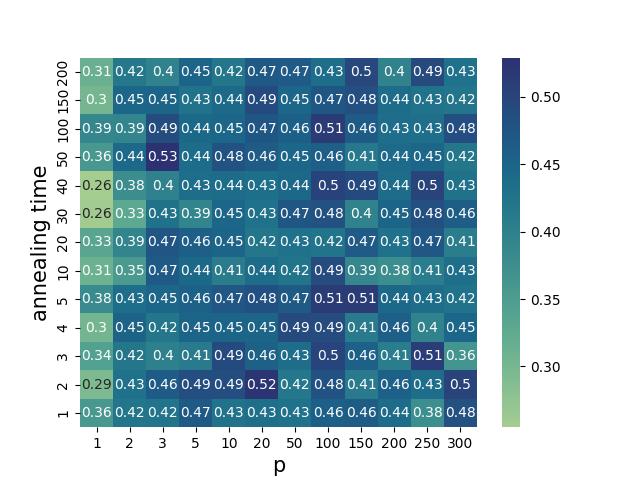}
	\includegraphics[trim=5mm 0 15mm 5mm, clip, width=0.495\textwidth]{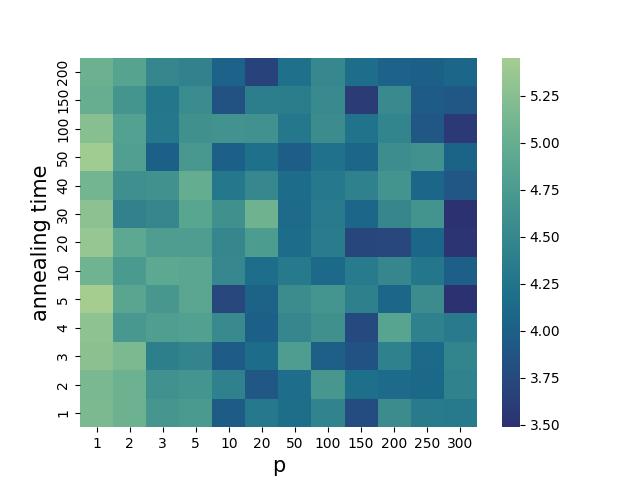}
	
	\includegraphics[trim=5mm 0 15mm 5mm, clip,width=0.495\textwidth]{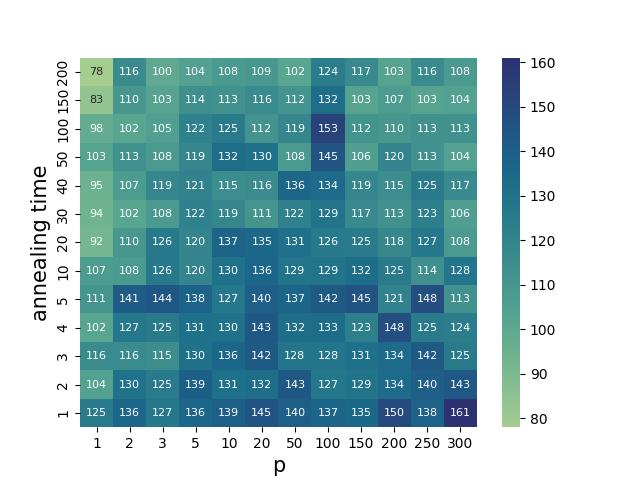}
	\includegraphics[trim=5mm 0 15mm 5mm, clip,width=0.495\textwidth]{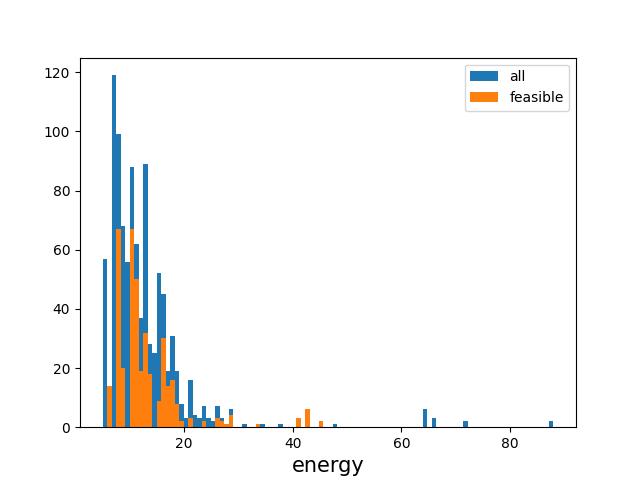}
	\caption{Feasibility rate (upper left), averaged Hamming distance (upper right), and (lower left) number of individual solutions as a function of anneal time ($\mu$s) and penalty factor $p$. (Lower right) energy distribution at $p=2$ and annealing time of 20$\mu s$. Values based on samples generated by D-Wave Advantage\texttrademark\ to solve our trivial ILP problem.}
	\label{fig:feas_rate_and_indep}
\end{figure}

In the brute force sampling case, we can describe the distribution of solutions upon the Hamming distance (\autoref{fig:dwave solutions ILP} left, middle row) by a cumulative distribution function,  
\begin{equation}
    CDF(d) = \frac{1}{2^{N_{\ve q}}}\sum_d \binom{N_{\ve q} }{d} \quad\forall\quad d \in \{0, 1, \ldots, N_{\ve q}\}\,,
\end{equation}
with $d$ representing possible Hamming distances for binary search vectors $\ve q$ of length $N_{\ve q}$. This relation can be used for benchmarking as it forms a fundamental boundary that only depends on the vector size.

In \autoref{fig:feas_rate_and_indep} (upper left panel) we show dependence of the feasibility ratio \eqref{eq:feasi_ratio}
as a function of anneal time and penalty factor, as well as the average hamming distance (upper right panel) as a function of the same parameters. 
There is seemingly little correlation between $p$ and the anneal time as long as $p\gtrsim10$.
However, these results suggest that increasing beyond $p\gtrsim 100$ is beneficial since in this region  solutions with lower Hamming distance are more likely. It is remarkable, that all feasibility ratios that are shown in \autoref{fig:feas_rate_and_indep} are significantly larger than the theoretical value of 18.75\% for the case that all possible solutions are considered.

We also encountered a number of individual feasible solution samples obtained in a single run whereby the solution vectors fulfil the ILP's constraints and differ from each other in at least one of its components, but are not necessarily optimal solutions. It can happen that some of these individual feasible solutions can share the same cost value.
The parameter dependence of the number of individual feasible solutions is presented in lower left panel of \autoref{fig:feas_rate_and_indep}. 
We find that short annealing times generate more individual solutions, however at the expense of reducing the low-energy solutions.
So the D-Wave quantum annealer finds more solutions with higher energies if shorter annealing profiles are applied. To no surprise, these correlations suggest that optimizing to longer anneal times will provide lower energy solutions. 
Similar findings are found for the other trivial ILPs listed in the Supplementary Material.
The relevant figures in this case are Figs.~S1-S3.

We point out that we find no correlations between the parameters \texttt{chain\_strength} and \texttt{annealing\_time}, suggesting that further optimization of the \texttt{chain\_strength} parameter is not possible.

\subsubsection{Improvements obtained by Machine Learning Approach}
\label{sec:ML improv}
Within a sample set, generated by D-Wave Advantage\texttrademark\, we have 110 independent solutions for our trivial ILP problem when using $p=2$ and an annealing time of 20 $\mu$s. This represents approximately 10\% of the possible feasible solutions.
To improve upon this, we train a NN on the correlations described above and then use the NN to generate more solutions.

\begin{figure}[h]
	\centering
	\includegraphics[trim=3mm 1mm 1mm 2mm, clip,width=0.49\textwidth]{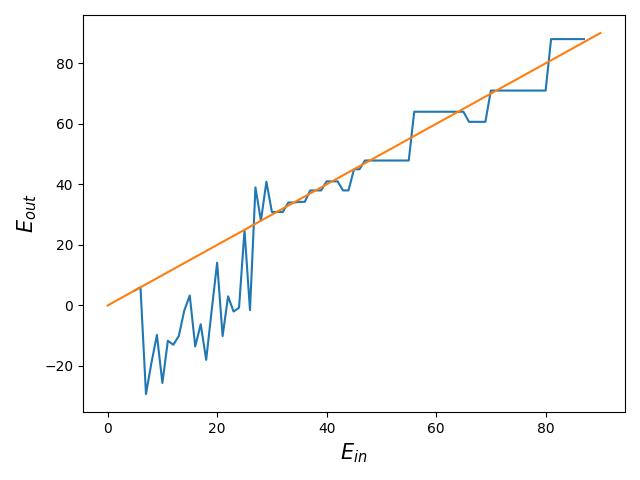}
	\includegraphics[trim=3mm 0 15mm 5mm, clip,width=0.49\textwidth]{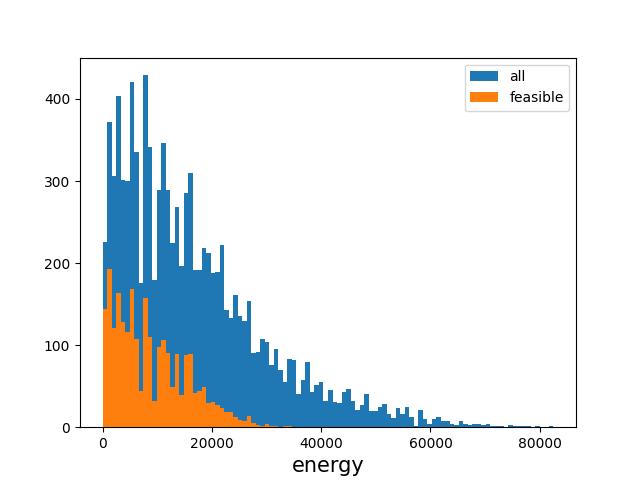}
	\caption{\label{fig:ML} Decision tree method to find more feasible solutions based on D-Wave data at $p=2$ and annealing time=20 $\mu s$.}
\end{figure}

In particular, we train our NN using the solution vector versus the energy and feasibility correlations obtained from D-Wave data.
With input of energy and feasibility, the decision tree regression predicts a new solution which has the corresponding input energy and feasibility. 
We note that our NN does not always provide new solutions whose output energy coincides with the same input energy. 
This is readily seen in the left plot in \autoref{fig:ML}, where the output energy $E_{out}$ is plotted as a function of input energy $E_{in}$.  
A one-to-one correspondence would provide a straight line with slope of unity, which is clearly not seen.  
However, the correlation between input and output energy as captured by our NN is still positive.
We find that the slope of this correlation depends on the $p$ value, whereby larger $p$ values provide a slope closer to unity. 
Qualitatively similar behavior is found for the ILPs listed in the Supplementary Material, as can be seen in Fig.~S5 of this document.

We expect the decision tree to recognize the feasibility condition, but predicted solutions of the NN are not always feasible.
As mentioned above, the NN predicts solution vectors whose energy ranges have some correlation with the input energies. 
This feature provides, in principle, an advantage over brute-force sampling since we can target solutions within a specific energy range using our NN, whereas such control via brute force sampling is not possible.
However, there isn't a complete one-to-one correspondence between input and output energy since approximately 20\% of the predicted solution vectors have components that are not binary but contain fractional numbers. 
In these cases we round the fraction to zero if the fractional number is smaller than 0.2, and to one if larger than 0.8. Between 0.2 and 0.8, we enumerate all possible combinations of 0 and 1, generating in these case new proposed solution vectors. 
We then perform another feasibility test on these NN solutions to filter out infeasible solutions.
The energy distribution of feasible versus infeasible solutions after this treatment is shown in \autoref{fig:ML} right.

\subsection{Three-node Network}
\begin{figure}[h]
	\centering
	\includegraphics[width=.8\textwidth]{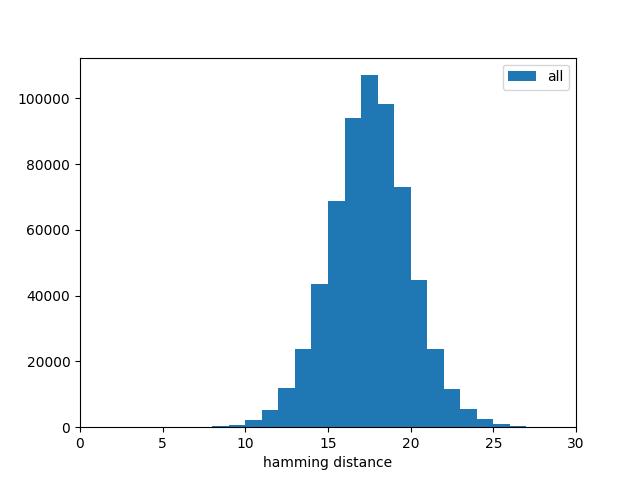}
	\caption{\label{fig:hamming_histo} Histogram of Hamming distance of the all solutions for 3-node problem. There are no feasible solutions.}
\end{figure}

\begin{figure}[h]
	\centering
	\includegraphics[trim=5mm 0 15mm 5mm, clip,width=0.495\textwidth]{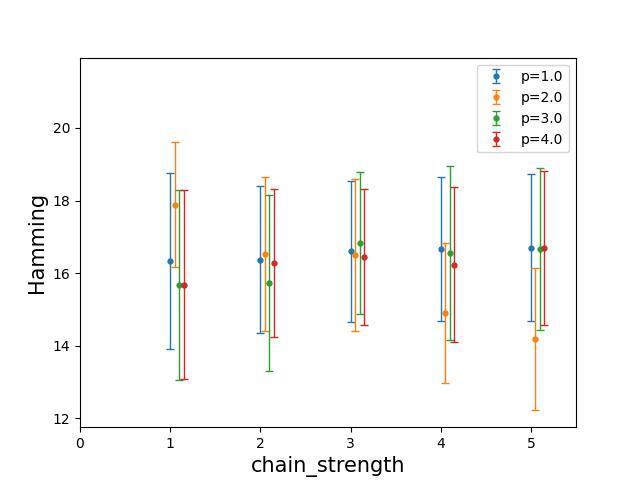}
	\includegraphics[trim=5mm 0 15mm 5mm, clip,width=0.495\textwidth]{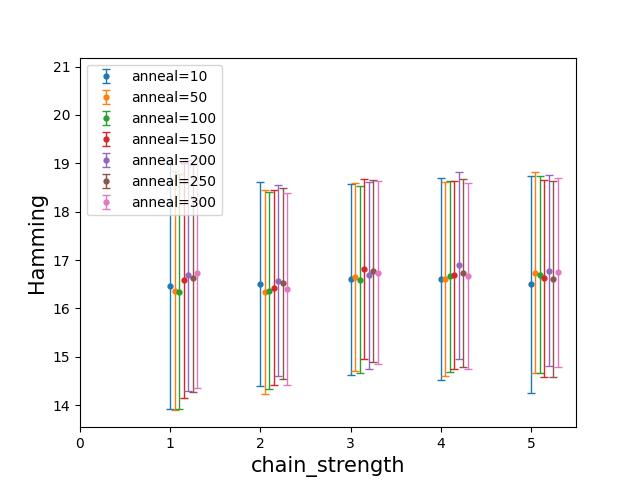}
	\caption{\label{fig:hamming} Hamming distance in dependence of penalty p and annealing parameters chain strength and annealing time. The Hamming distance is obtained by comparing with the best solution obtained by CPLEX.}
\end{figure}

We now turn our attention to the 3-node problem, which represents the smallest, non-trivial system of wide-area networks. 
Here we use CPLEX to obtain the optimal solution vector, from which we make comparisons with D-Wave solution vectors.
The distribution of D-Wave solutions as a function of Hamming distance to the optimal solution is given in \autoref{fig:hamming_histo}.
Note that in this case the optimal solution is not captured by D-Wave.  
In fact, D-Wave cannot find any feasible solutions within a set of 600 000+ samples. As remark, the entire search space for this case is $2^{63}$.

When we investigate inter-parameter correlations, we find little to no correlations between the Hamming distance, \texttt{chain\_strength}, \texttt{anneal\_time}, and penalty factor $p$.
This is demonstrated by the nearly flat dependence of the data in \autoref{fig:hamming}. 
This lack of correlation prevents us from obtaining optimized run parameters for this system, and unfortunately suggests that larger node problems will become just as difficult, if not more difficult, to optimize.

These findings already hint at the difficulties we encounter when applying an NN to this system, as we describe in the following section.
But we nonetheless train a DT network on the energy and Hamming distance to optimal solution, exactly as described in \autoref{sec:ML improv}.

\section{Discussion}
\label{sec:conclusion}

\subsection{Interpretation of Findings}
\begin{figure}[b]
	\centering
	\includegraphics[width=.8\textwidth]{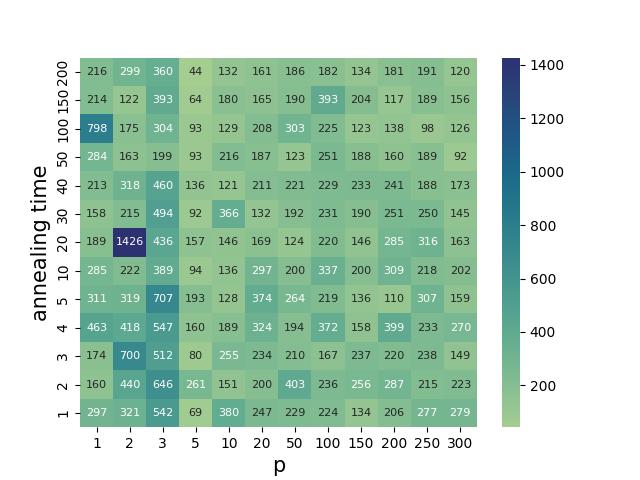}
	\caption{\label{fig:new_indep_sol} The number of new independent feasible solutions found by decision tree for the ILP problem.}
\end{figure}

\begin{figure}[h]
	\centering
	\includegraphics[width=.8\textwidth]{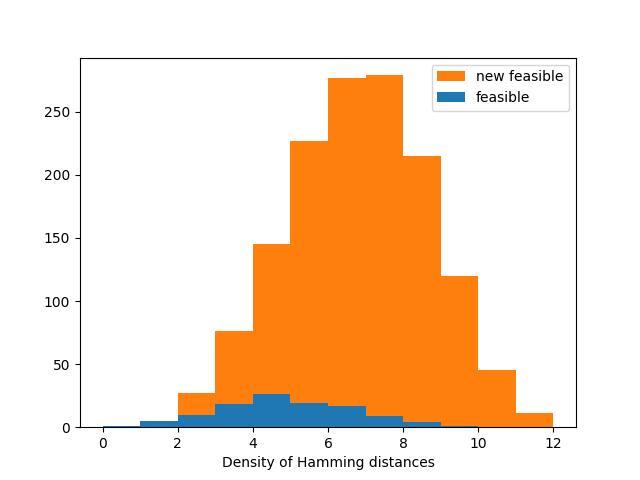}
	\caption{\label{fig:hamming_new} Histogram of Hamming distance of the feasible solution obtained by D-Wave (blue) and the new feasible solutions obtained by decision tree method for the ILP problem.}
\end{figure}

The total number of feasible solutions of the trivial ILP problem is 1536. 
As previously mentioned, D-Wave finds a little less than 10\% of these solutions, but with our NN we can fully ascertain the full solution space distribution (compare the lower right panel of \autoref{fig:feas_rate_and_indep} with that of \autoref{fig:ML} and see also Figs.~S2 and S5 of our Supplementary Material). 
More concretely, we provide the exact number of addition feasible solutions found with our NN as a function of input parameters \texttt{annealing time} and penalty factor $p$ in \autoref{fig:new_indep_sol} (see Fig.~S4 for our other trivial ILPs). 
This means that, for our simple ILP problem, the decision tree after round off treatment provided 1426 new independent feasible solutions.
The distribution of new solutions as a function of Hamming distance provided by our ML technique is given in \autoref{fig:hamming_new}.
So combining our NN results with D-Wave's, all possible 1536 solutions were found.
Thus our hybrid classical (ML)/quantum (D-Wave) method allowed us to fully map out the full solution space.
We note that our NN is not generalizable to all trivial ILPs, but is unique for each ILP.
This is because the solution vector space generated by D-wave is specific for each ILP, and so each NN is trained with this specific solution vector layout.
Within our formalism a `master' NN for all trivial ILPs is not possible.

We now discuss our 3-node problem.  
Note that in this case, D-Wave could not find the optimal solution provided by CPLEX, despite our system parameter investigations mentioned in the previous section. It is not viable to assume that feasible solutions can be found by luck or random guessing. The probability to find the optimal (minimal and feasible) solution is $1/2^{63}\sim10^{-19}$ in our case. The fact that we could not find feasible solutions within a set of \SI{600000}{samples} indicates that feasible solutions are very rare. This was already observed in our previous study \cite{Witt:2022lsx}. There, we were not able to find any feasible solutions for some of the test sets and in other cases around 0.2 to 11 per million samples. There are some possible hints for why this is the case here. First of all, the entire solution set contains only a small portion of feasible solutions that fulfill the ILP. Further, the annealer minimizes the energy of the QUBO Hamiltonian. As it is possible that the lowest energy state can be obtained with various solution vectors someone could find also a energy-wise optimized vector that does not fulfill the ILP. Furthermore, hardware imperfections like noise or limited detection resolution can cause this undesirable behavior.

At this point, critical voices could rate the annealer as an expensive random sampler. But this is not the case as we were able to show that trivial ILP problems are definitely solvable with D-Wave. In these cases, we explicitly used less samples than the solution space's size to avoid an oversampling---somebody could also solve small problems by oversampling even if the sampler is neither a random guess sampler where each solution is equal probable or a minimizing sampler like the quantum annealer. Thus, D-Wave performs better than a random sampler. Clearly the solvability is not the same for the network problem case, and this may be due to a) a higher connected QUBO and longer chains of qubits that represent a logical qubit, which cause chain breaks in the quantum annealing hardware to be more likely, b) numbers in the QUBO matrix have a higher differing range that may be not represented in the hardware well-enough, and of course c) other issues that are beyond our knowledge.

\begin{figure}[h]
	\centering
	\includegraphics[trim=5mm 0 15mm 5mm, clip,width=0.495\textwidth]{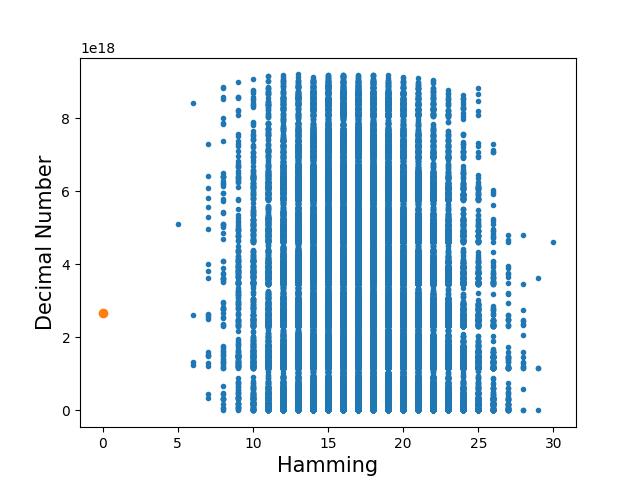}
	\includegraphics[trim=5mm 0 15mm 5mm, clip,width=0.495\textwidth]{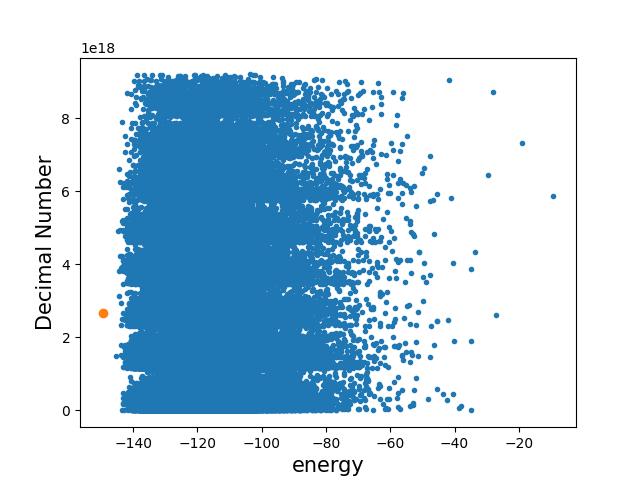}
	\caption{\label{fig:decimal} Comparing the data distribution and the best solution. We convert the solution vector to decimal number to plot.}
\end{figure}

As part of an approach for improvements, we trained our NN for the 3-node problem with the distributions that we generated from our correlation studies in a comparable way as it was done in the simple ILP problem.
Once trained, we found, however, that the NN was unsuccessful in finding any new feasible solutions, let alone the optimal solution.
We attribute this to the fact that our D-Wave data distribution of energies (which is used to train the NN) does not cover the energy region of the optimal solution.
In fact, as shown in \autoref{fig:decimal}, the distribution of D-Wave solutions is far from the optimal solution. 
Our NN could therefore not generalize sufficiently to lower energy solutions.
Compounding the issue is the fact that the distribution of D-Wave solutions contained no feasible solutions, and this in turn limited what the NN could `learn'.  
Thus our hybrid (ML)/quantum (D-Wave) method failed to produce any new solutions for our 3-node problem.

An obvious question to raise is whether another choice of NN is better suited for our 3-node problem.
As we discussed in~\autoref{sec:NN}, one of the main advantages that motivated our choice of the decision tree NN is admittedly its ease of use, interpretability, and implementation.
However, because of its potential lack of expressivity, one could argue that another choice of NN, e.g. convolutional or recurrent, might lead to better results.  
This indeed may be the case, and at the least warrants further research.
We point out, however, that regardless of the NN architecture, our formalism requires that there exist correlations between hyper-parameters and the resulting D-Wave solutions vectors.
It is these correlations that are `learned' by the NN.
Since we found no such correlations in our 3-node problem, we suspect that any other type of NN will have similar difficulties as those encountered by our decision tree NN.

\subsection{Outlook on Further Improvements}
Still there may be ways to improve the situation.  
Our studies to date have only varied the annealing profile.
Instead, one may perform reverse annealing, where the annealing is run `backwards' from a starting classical solution, allowing for exploration of the energy landscape around the classical solution.  
We are actively investigating this procedure. Reverse annealing may be also applicable to set initial states as shown in \cite{rev_anneal}. Thus, expected solutions or solutions that are close to an expected solution can be set as start value for the annealing process. If the optimizer is applied frequently---a typical situation in network optimization---the last obtained solution can be used for the initialization of the next run as new optimal network configurations might be close to the last configuration.

Annealing parameter like annealing schedules and various embeddings for our problems can be studied more detailed like in the study of \cite{Willsch_2022}. The authors of \cite{Willsch_2022} discovered an increase in the success rate for proper settings in the annealing schedule. In our case we observed a more or less constant success rate, especially for the 3-node network problem. Apart from that it may be valuable to study our approach on a larger set of similar problems to get a more general perspective. Unfortunately, we had to restrict our study on a single problem instance as the amount of feasible solutions for our problem is very rare and large sampling sets are required for the analyzes. Besides, thermalization within the annealing process can be studied as well, see \cite{term_annealing_2013}.
	
\begin{figure}[h]
	\centering
	\includegraphics[width=.495\textwidth]{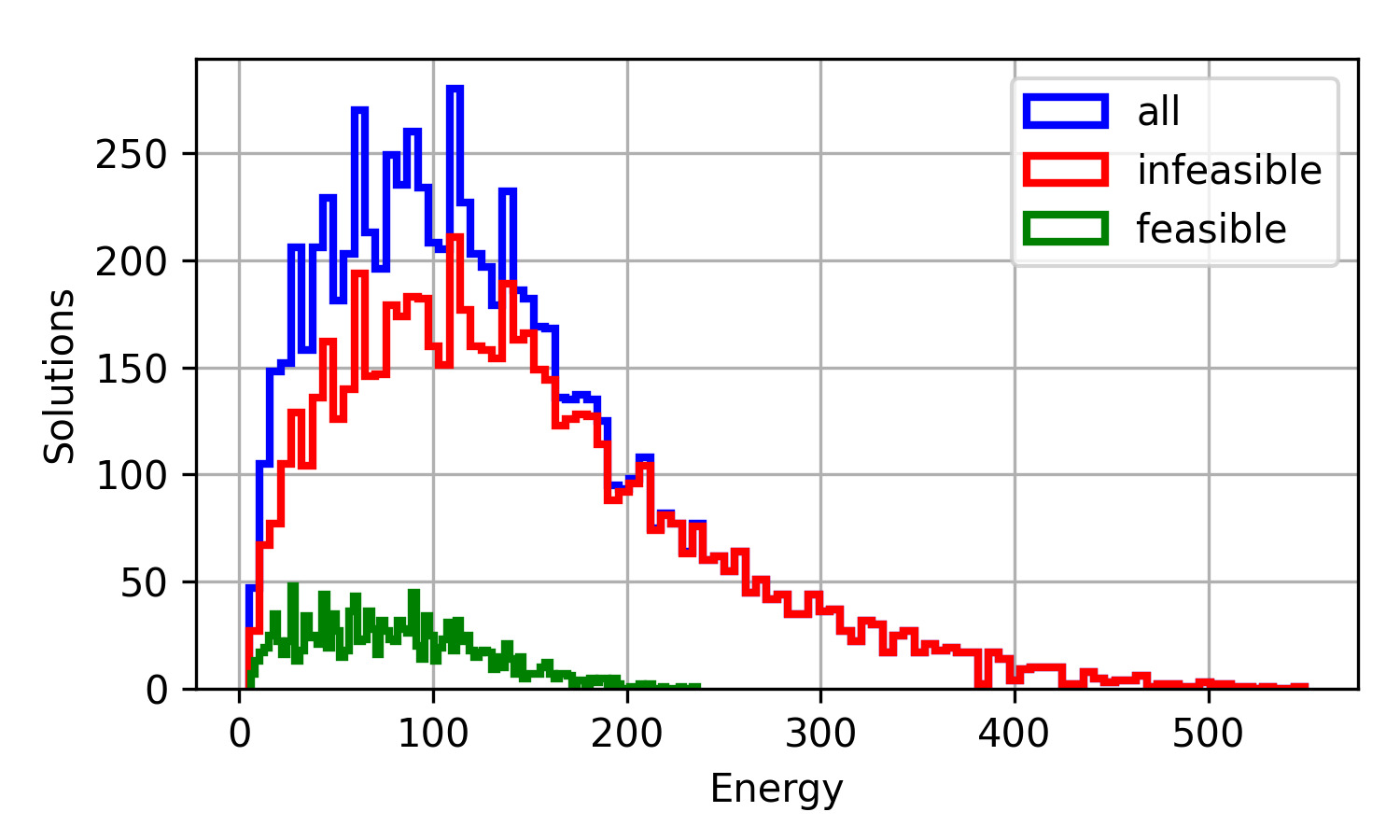} 
	\includegraphics[width=.495\textwidth]{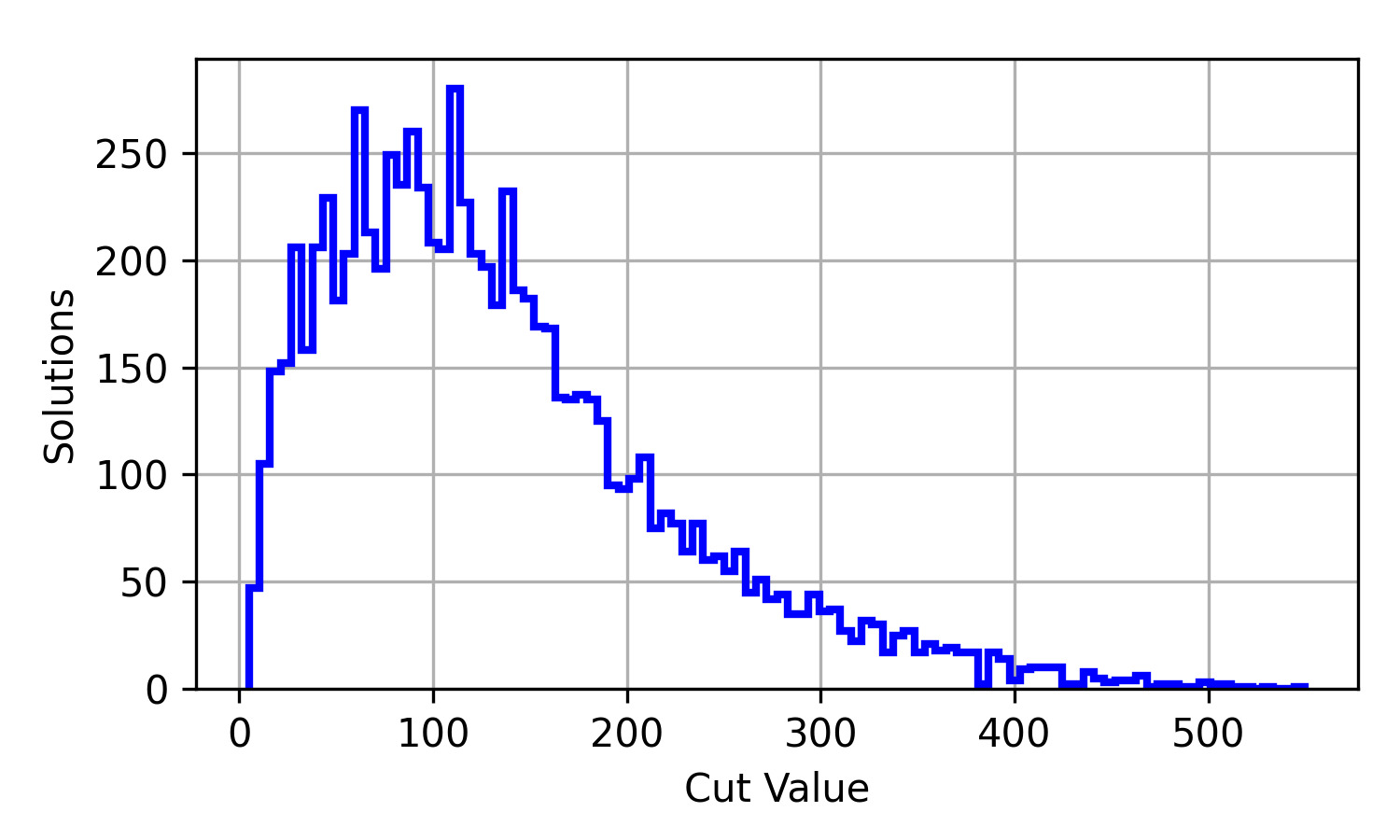}
	\caption{Evaluation of the trivial ILP problem in representation as Sherrington-Kirkpatrick (SK) graph. (Left) distribution of energy for SK graph's Hamiltonian if all possible solutions are considered, (right) distribution of all cut values in SK graph.} 
	\label{fig:sk_graph}
\end{figure}

Furthermore, since the size of the problem that is embedded on the quantum annealer plays a crucial role for its solvability, methods for efficient embedding or problem reduction should be incorporated within future studies. We point out that the work of \cite{fasthare} seems promising in reducing the demands on the number of physical qubits. Here the authors introduced a fast Hamiltonian reduction algorithm (FastHare) that defines non-separable groups of qubits, i.e. qubits that obtain the same value in optimal solutions, and performed a reduction by merging non-separable groups into single qubits. This could be done within a worst case time complexity of $\mathcal{O}(\alpha n^2)$ with a user-defined parameter $\alpha$. The authors of \cite{fasthare} showed in a benchmark that their algorithm is capable of saving 62\% of physical qubits on average within a processing time of 0.3 seconds, outperforming the roof duality--the reduction used within the D-Wave's software development kit SDK. 
We reviewed parts of their work. In particular, we mapped our trivial ILP problem to a so-called Sherrington-Kirkpatrick (SK) graph. We further evaluated all cut values within this graph. The results \autoref{fig:sk_graph} show that the cut values in the SK graph correspond to the energy values of QUBO or Ising problem solution vectors. As the Hamiltonian reduction is based on graph compression on basis of minimal cuts, we expect that the proposed algorithm \cite{fasthare} can improve the situation, as a reduced Hamiltonian might be better solvable on the D-Wave quantum annealer.

Unfortunately, we were not able to fully implement and apply this sophisticated algorithm as we struggled at the following point. The algorithm applies a min-cut algorithm on the SK graph to detect non-separable qubit groups. Originally, we though that a standard min-cut algorithm could be applied here. Unfortunately, the for us available min-cut algorithms can be only applied in graphs with positive-weighted edges. But, due to the nature of QUBO, Ising or SK Hamiltonians, the edges in a SK graph may have negative-valued edge weights. This issue was not addressed in their work \cite{fasthare}. However, it remains unsure, if the fast Hamiltonian Reduction (FastHare) algorithm can improve the solvability of our ILPs with D-Wave's annealer as the authors used randomly generated graph structures in their evaluation, i.e., the graphs are weakly connected and as such well-suited for graph compression.

Beside the ILP to QUBO mapping formalism that was described in \cite{Chang:2020iwh} and \cite{Witt:2022lsx}, someone could model the problem in a differing way. One possibility is the introduction of constraint-specific penalty factors, that create new degrees of freedom usable for problem-specific optimization of the algorithm. It can be achieved by the use of a penalty vector $	\ve{p}^\top=[p_1,p_2,\ldots,p_m] $
and a corresponding penalty matrix $\ve{P}=\ve{I}\ve{p}$ inside the formulations. The QUBO Hamiltonian and thus the objective to be optimized is then
\def\Qxx{\ve{Q}_{xx}}
\def\Qxs{\ve{Q}_{xs}}
\def\Qsx{\ve{Q}_{sx}}
\def\Qss{\ve{Q}_{ss}}
\def\Zx{\ve{Z}_{x}}
\def\Zs{\ve{Z}_{s}}
\begin{equation}
	\mathcal{H}(q)=\ve{q}^\top\ve{Qq}+C \rightarrow\text{min}, \quad\text{with}\quad
	\ve{Q}=
	\begin{bmatrix}
		\Qxx & \Qxs \\
		\Qsx & \Qss \\
	\end{bmatrix}
	,\quad C= \ve{b}^\top\ve{Pb}\quad\text{and}\quad\nonumber
\end{equation}\vspace{-0.5cm}
\begin{eqnarray}
	\Qxx&\hspace{-2mm}=\hspace{-2mm}&\Zx^\top \ve A^\top \ve P \ve A \Zx + \diag\left\{\left(2\ve{b}^\top \ve P\ve A +\ve{c}^\top\right) \Zx    \right\}\,,
	\nonumber\\
	\Qxs&\hspace{-2mm}=\hspace{-2mm}&\Qsx^\top=\Zx^\top \ve A^\top \ve P \Zs\,,
	\nonumber\\
	\Qss&\hspace{-2mm}=\hspace{-2mm}&\Zs^\top \ve P \Zs+2\diag\left\{\Zs^\top \ve P \ve{b} \right\}\,.
	\label{eq:qubo_matrix}
\end{eqnarray}
This extends the ILP to QUBO mapping formalism to a generalized form. Required details could be found in \cite{Witt:2022lsx}, Sec.~III-D. 

\subsection{Outcome}
Our work can be summarized as follows. The approach aims to solve ILPs with a quantum annealing attempt . We tried to find optimal annealing parameters and discovered weak correlations between annealing parameters and success rates in the 3-node network case. Further, a decision tree ML approach was applied to increase the rate of feasible ILP solutions. We realized that further improvements are needed to overcome remaining hurdles and discussed some attempts therefore. Even as the results for the 3-node problem are not fully satisfying, we are able to show with less complicated ILP problems that the approach works in principle. Thus, we expect that the approach can be extended in a way that larger problem instances are also solvable.

Finally, fast ILP-solving methods can have a significant impact on systems that should be optimized in real time. As an example, a novel mode of real-time network operation in wide-area networks is studied in \cite{Witt_HPSR}. Here, similar ILPs are used to define a frequently applied network optimization.

\section*{Conflict of Interest Statement}
The authors declare that the research was conducted in the absence of any commercial or financial relationships that could be construed as a potential conflict of interest.

\section*{Author Contributions}
\textbf{A.W.:} Formal analysis, Investigation, Methodology, Resources, Software, Visualization, Writing -- original draft.
\textbf{J.K.:} Formal analysis, Investigation, Methodology, Resources, Software, Visualization, Writing -- review \& editing.
\textbf{C.K.:} Data curation, Writing -- review \& editing.
\textbf{T.L.: } Writing -- original draft,
Project administration

\section*{Funding}
A.W. was supported by the German Federal Ministry of Education and Research (Project ID 16 KIS 1312) which is partly funding the work that has been performed in the framework of the CELTIC-NEXT EUREKA project AI-NET ANTILLAS (Project ID C2019/3-3).

J.K. and T.L. were supported by the Deutsche Forschungsgemeinschaft (DFG, German Research Foundation) through the funds provided to the Sino-German Collaborative Research Center TRR110 "Symmetries and the Emergence of Structure in QCD" (DFG Project-ID 196253076 - TRR 110).

The Jülich Supercomputing Centre funded this project by providing computing time through the Jülich UNified Infrastructure for Quantum computing (JUNIQ) on the D-Wave Advantage\texttrademark\ quantum system.

This publication was funded by the German Research Foundation (DFG) grant "Open Access Publication Funding / 2023-2024 / University of Stuttgart" (512689491).

\section*{Acknowledgments}
The authors gratefully acknowledge the Jülich Supercomputing Centre for funding this project by providing computing time through the Jülich UNified Infrastructure for Quantum computing (JUNIQ) on the D-Wave Advantage\texttrademark\ quantum system. They further acknowledge the project funding received.

\bibliographystyle{Frontiers-Harvard} 
\bibliography{literature}

\clearpage
\iflatexml
\section*{Supplementary Material}
\else\fi


\section*{Supplementary material}

# *Supplementary Material*

## 1 SUPPLEMENTARY TABLES AND FIGURES

### 1.1 Tables

**Table S1.** Definition of various trivial integer linear programs for benchmarking and solutions obtainable by classical solver.

| ILP variant | $\mathbf{A}$ | $\mathbf{b}$ | $\mathbf{c}$ | bits to represent values in $\mathbf{x}$ | bits to represent values in $\mathbf{s}$ | solution | cost value |
|---|---|---|---|---|---|---|---|
| 1 | $\begin{bmatrix} -1/3 & -1 \\ -3 & -1 \\ 0 & 1 \end{bmatrix}$ | $\begin{bmatrix} -2 \\ -6 \\ 2 \end{bmatrix}$ | $\begin{bmatrix} 1 \\ 3 \end{bmatrix}$ | 2 | 3 | $\begin{bmatrix} 3 \\ 1 \end{bmatrix}$ | 6 |
| 2 | $\begin{bmatrix} -2 & -2 \\ -1 & -4 \\ 1 & 0 \end{bmatrix}$ | $\begin{bmatrix} -3 \\ -5 \\ 2 \end{bmatrix}$ | $\begin{bmatrix} 2 \\ 1 \end{bmatrix}$ | 2 | 3 | $\begin{bmatrix} 0 \\ 2 \end{bmatrix}$ | 2 |
| 3 | $\begin{bmatrix} -2 & -2 \\ -3 & -4 \\ 1 & -1 \end{bmatrix}$ | $\begin{bmatrix} -1 \\ -6 \\ 0 \end{bmatrix}$ | $\begin{bmatrix} 1 \\ 2 \end{bmatrix}$ | 2 | 3 | $\begin{bmatrix} 1 \\ 1 \end{bmatrix}$ | 3 |
| 4 | $\begin{bmatrix} -1 & -2 & -3 \\ -3 & -4 & -1 \\ -1 & -1 & 0 \end{bmatrix}$ | $\begin{bmatrix} -5 \\ -5 \\ -3 \end{bmatrix}$ | $\begin{bmatrix} 1 \\ 2 \\ 1 \end{bmatrix}$ | 2 | 3 | $\begin{bmatrix} 3 \\ 0 \\ 1 \end{bmatrix}$ | 4 |
| 5 | $\begin{bmatrix} -1/3 & -1 \\ -3 & -1 \\ -1 & 1 \end{bmatrix}$ | $\begin{bmatrix} 2 \\ 6 \\ 0 \end{bmatrix}$ | $\begin{bmatrix} 3 \\ -3 \end{bmatrix}$ | 2 | 3 | $\begin{bmatrix} 0 \\ 0 \end{bmatrix}$ | 0 |

## 1.2 Figures

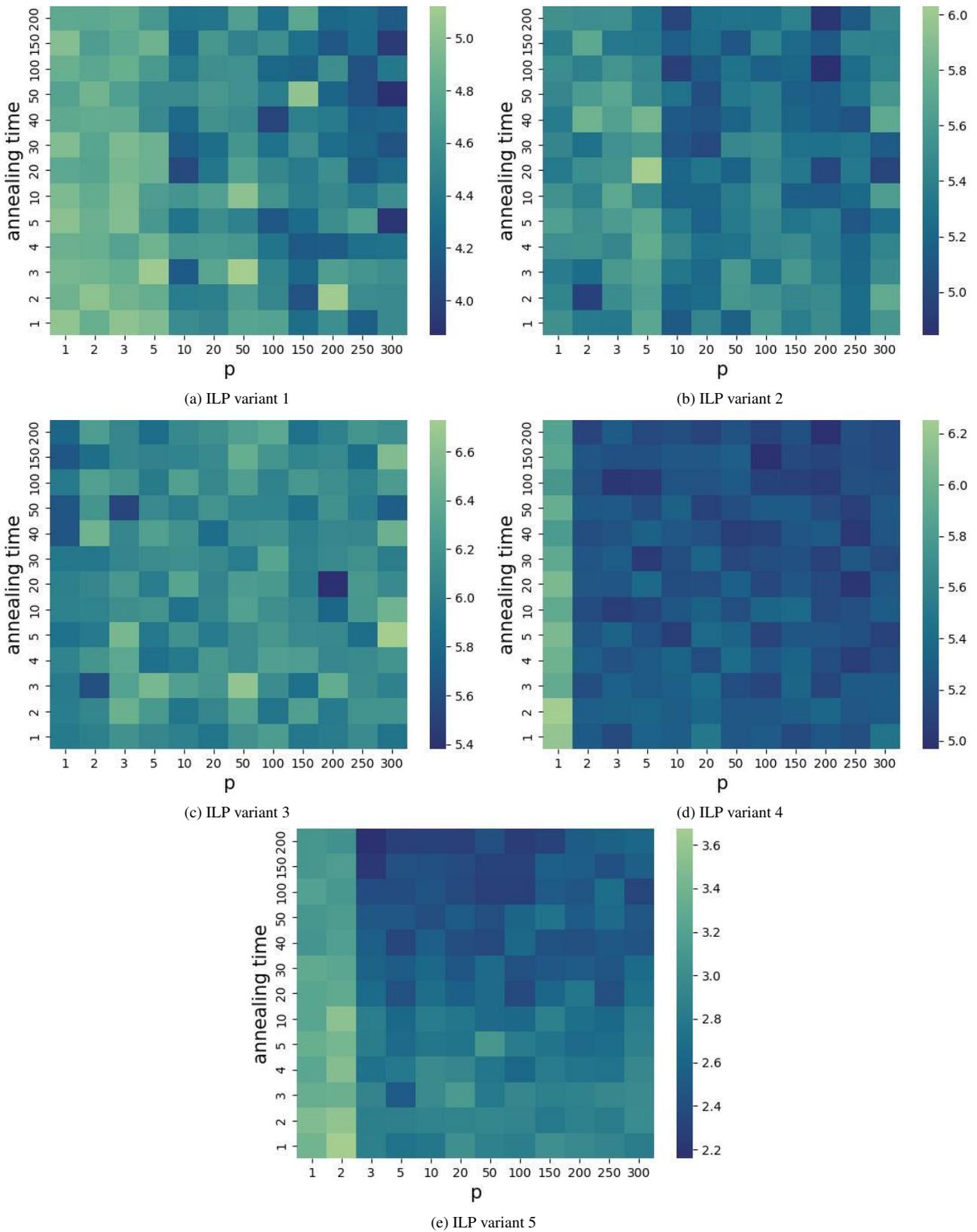

Figure S1: Average values of the Hamming distances between the best known solution and solutions obtained by the D-Wave Advantage™ quantum annealer for various trivial ILP problems that are defined according to Table S1. Varied parameters are the QUBO-specific penalty $p$ and the annealing time.

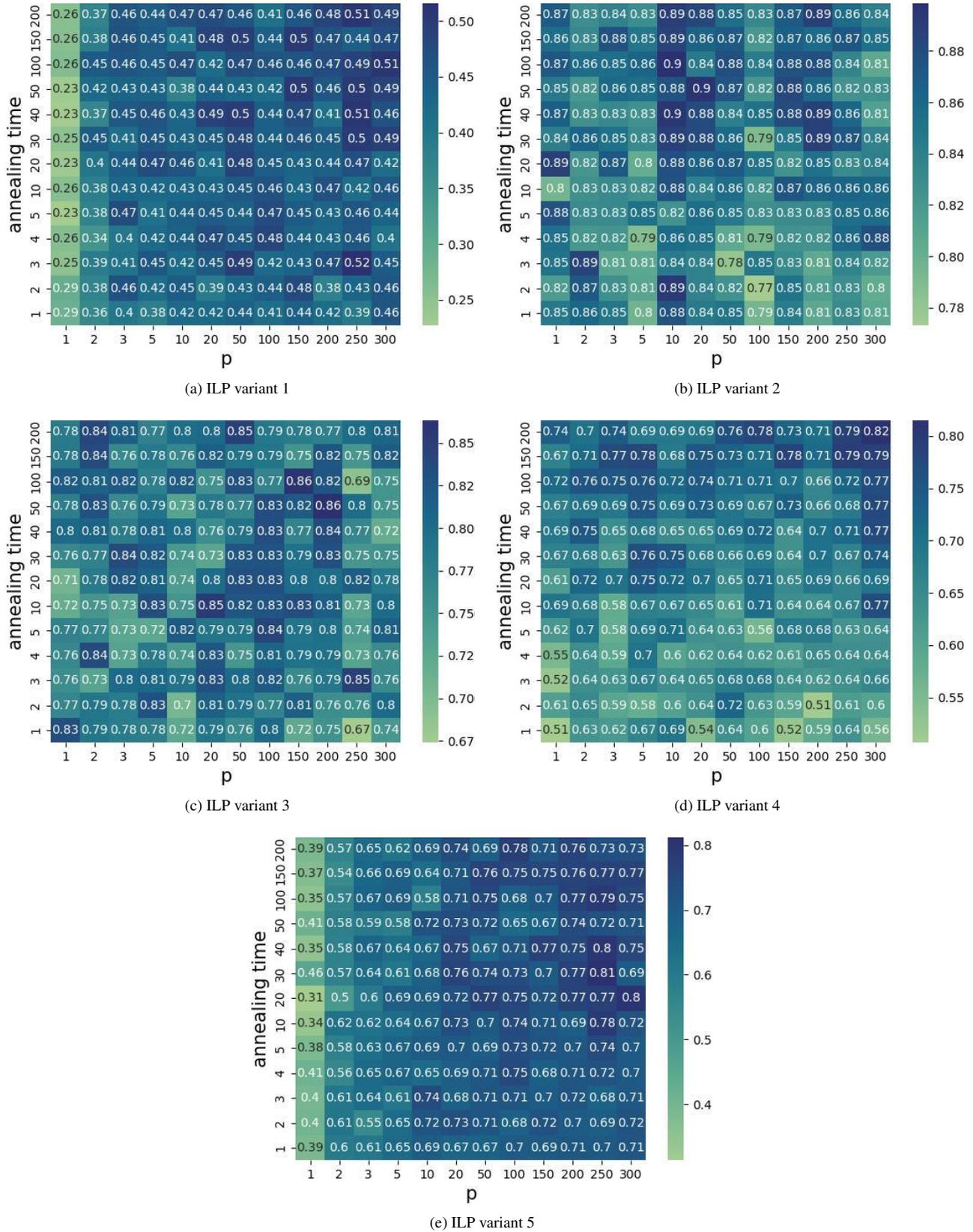

Figure S2: Rate of feasible solutions obtained by the D-Wave Advantage™ quantum annealer for various trivial ILP problems that are defined according to Table S1. Varied parameters are the QUBO-specific penalty $p$ and the annealing time.

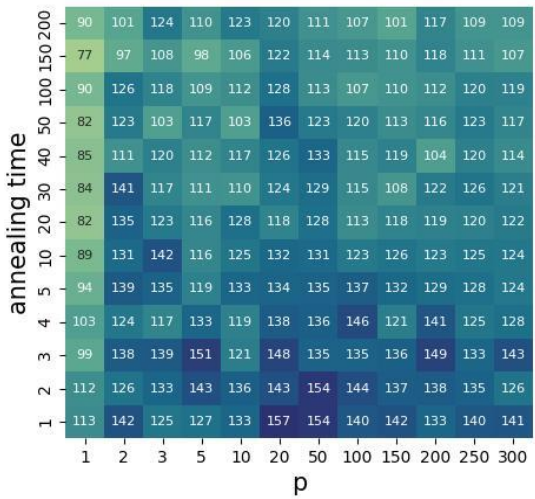
(a) ILP variant 1

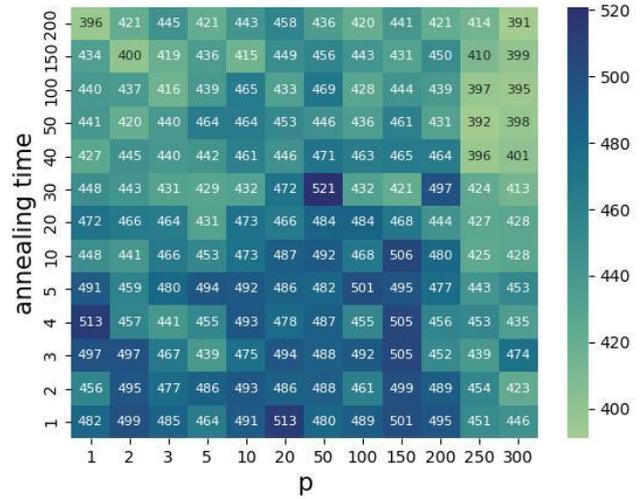
(b) ILP variant 2

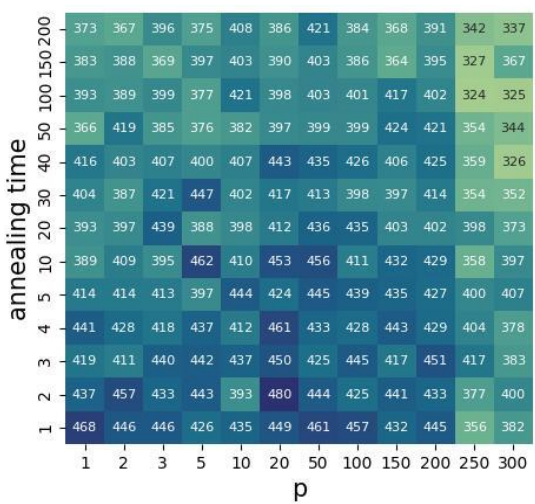
(c) ILP variant 3

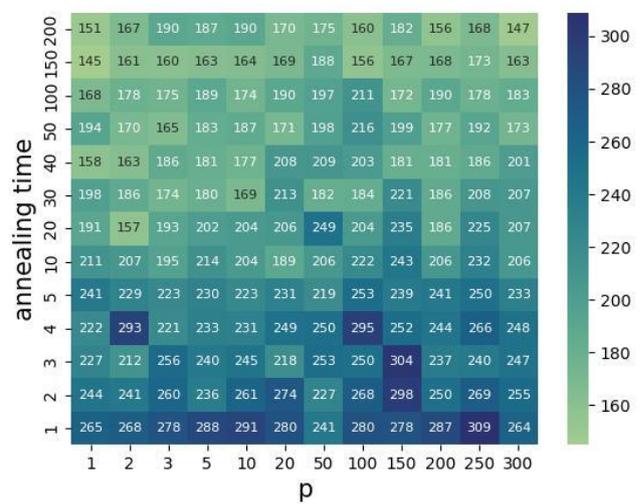
(d) ILP variant 4

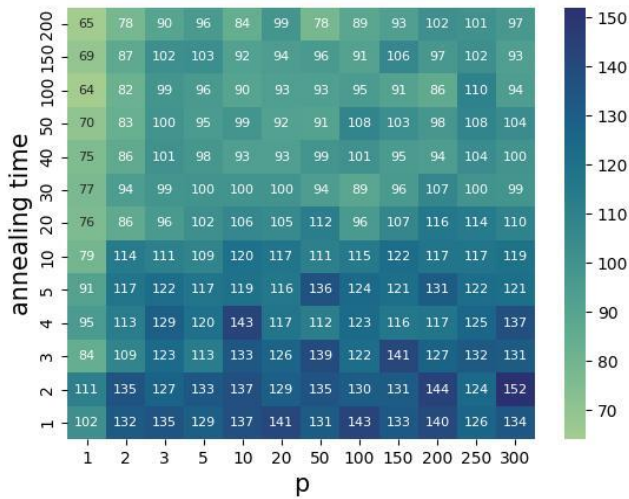
(e) ILP variant 5

Figure S3: Amount of individual feasible solutions obtained by the D-Wave Advantage™ quantum annealer for various trivial ILP problems that are defined according to Table S1. Varied parameters are the QUBO-specific penalty $p$ and the annealing time.

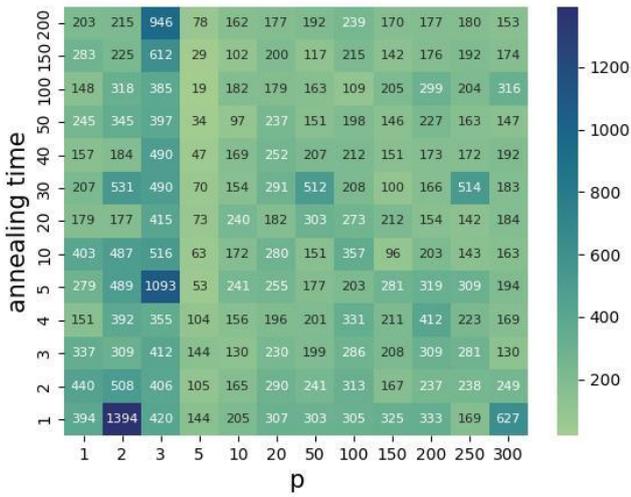

(a) ILP variant 1

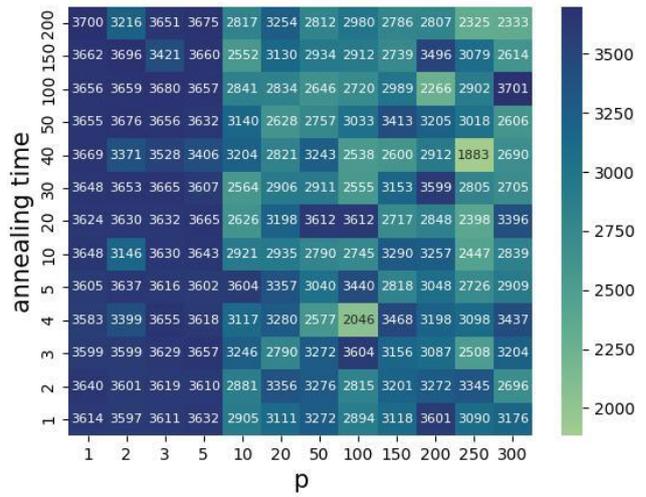

(b) ILP variant 2

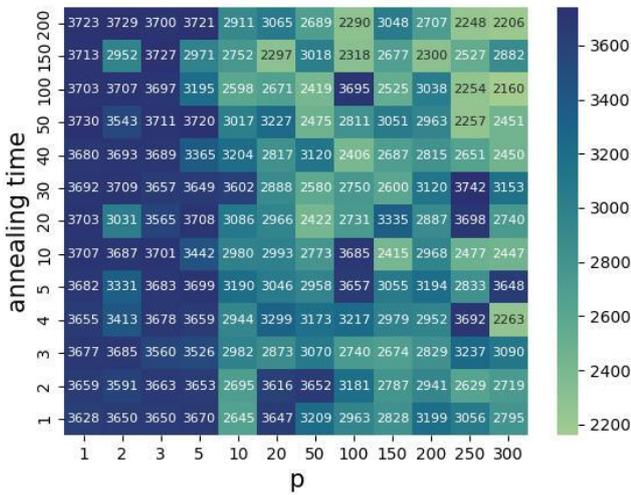

(c) ILP variant 3

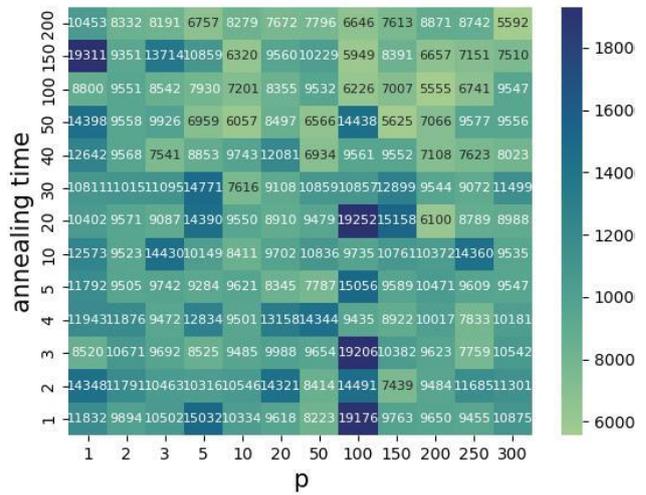

(d) ILP variant 4

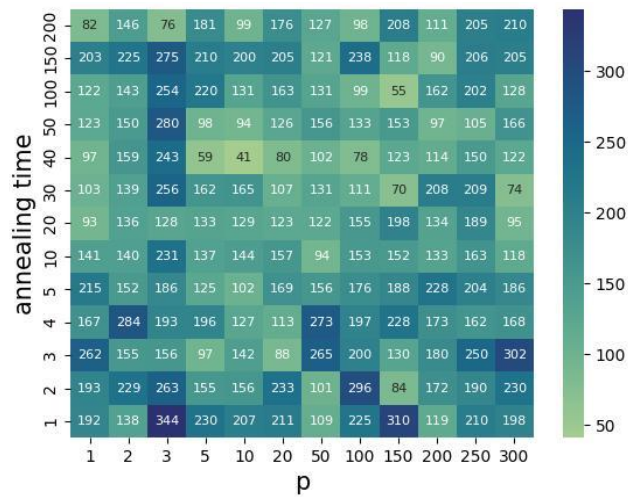

(e) ILP variant 5

Figure S4: Amount of new feasible solutions obtained by a neural network based on the decision tree method. Results are shown for various trivial ILP problems that are defined according to Table S1. Varied parameters are the QUBO-specific penalty $p$ and the annealing time.

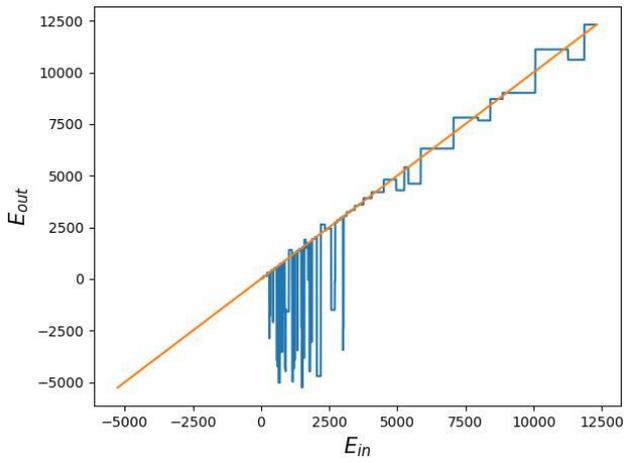

(a) ILP variant 1

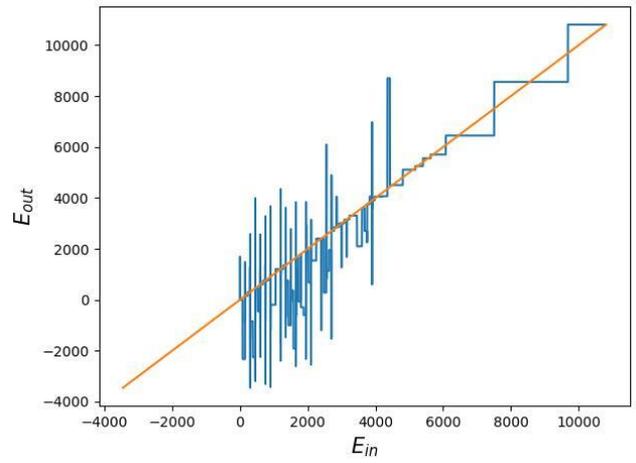

(b) ILP variant 2

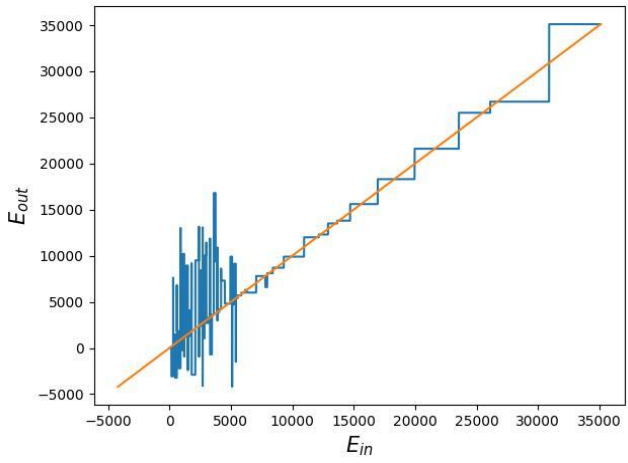

(c) ILP variant 3

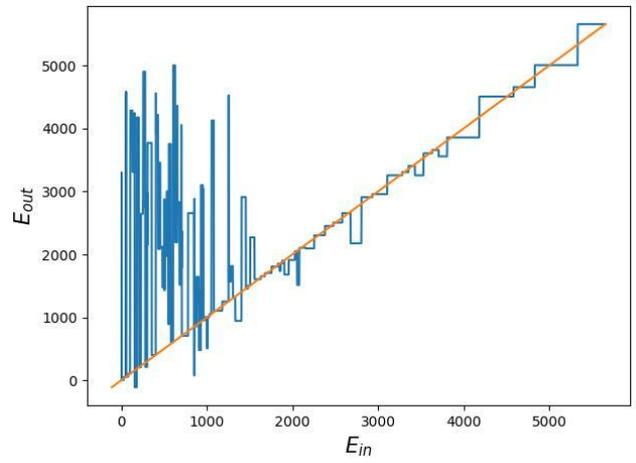

(d) ILP variant 4

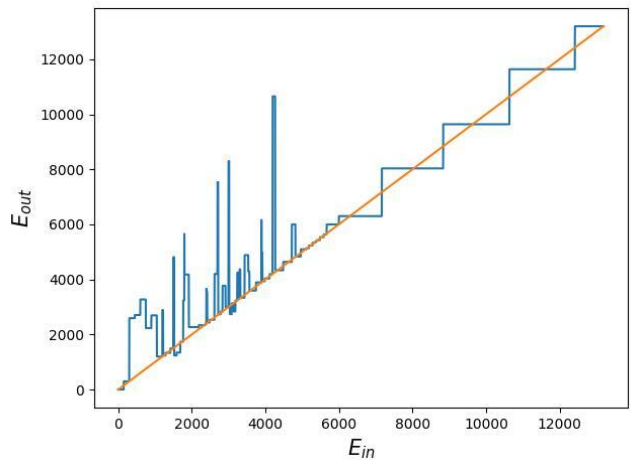

(e) ILP variant 5

Figure S5: Coincidence of energy values for input and output samples obtained by the decision tree method. Results are shown for various trivial ILP problems that are defined according to Table S1.



\end{document}